\RequirePackage{rotating}
\documentclass[]{aastex62}
\usepackage[utf8]{inputenc}
\usepackage[normalem]{ulem}
\usepackage{amsfonts,amsmath,amssymb}

\usepackage{xspace}
\usepackage{natbib}
\usepackage{verbatim}
\usepackage{graphicx,epstopdf}
\usepackage{fancyref}
\usepackage{hyperref}
\usepackage{breakurl}
\usepackage{amsmath}
\usepackage{graphicx} 
\usepackage{epstopdf}
\usepackage{graphicx}
\usepackage{appendix}
\usepackage{rotfloat}

\usepackage{amsmath}
\usepackage{graphicx}
\usepackage{natbib}
\usepackage{color}
\usepackage{rotating}
\usepackage{float}
\usepackage{booktabs}
\usepackage{graphicx}
\usepackage{enumitem}

\newcommand{\ha}{H$\alpha$}

\newcommand{\hb}{H$\beta$}
\newcommand{\mgii}{Mg\,{\sc ii}}
\newcommand{\civ}{C\,{\sc iv}}

\newcommand{\kmps}{$\rm km\,s^{-1}$}

\accepted{by ApJS}

\begin{document}
\title{The Large Sky Area Multi-object Fiber Spectroscopic Telescope (LAMOST) Quasar Survey: Quasar Properties from Data Release Six to Nine}

\author[0000-0002-8402-3722]{Jun-Jie Jin}
\affiliation{Department of Astronomy, School of Physics, Peking University, Beijing 100871, People's Republic of China; E-mail: jjjin@pku.edu.cn; wuxb@pku.edu.cn}
\affiliation{Kavli Institute for Astronomy and Astrophysics, Peking University, Beijing 100871, People's Republic of China}
\author[0000-0002-7350-6913]{Xue-Bing Wu}
\affiliation{Department of Astronomy, School of Physics, Peking University, Beijing 100871, People's Republic of China; E-mail: jjjin@pku.edu.cn; wuxb@pku.edu.cn}
\affiliation{Kavli Institute for Astronomy and Astrophysics, Peking University, Beijing 100871, People's Republic of China}
\author[0000-0002-0759-0504]{Yuming Fu}
\affiliation{Department of Astronomy, School of Physics, Peking University, Beijing 100871, People's Republic of China; E-mail: jjjin@pku.edu.cn; wuxb@pku.edu.cn}
\affiliation{Kavli Institute for Astronomy and Astrophysics, Peking University, Beijing 100871, People's Republic of China}
\author[0000-0002-9728-1552]{Su Yao}
\affiliation{Max-Planck-Institut f\"ur Radioastronomie, Auf dem H{\"u}gel 69, 53121 Bonn, Germany}
\author{Yan-Li Ai}
\affiliation{College of Engineering Physics, Shenzhen Technology University, Shenzhen 518118, China}
\affiliation{Shenzhen Key Laboratory of Ultraintense Laser and Advanced Material Technology, Shenzhen 518118, China}
\author{Xiao-Tong Feng}
\affiliation{Department of Astronomy, School of Physics, Peking University, Beijing 100871, People's Republic of China; E-mail: jjjin@pku.edu.cn; wuxb@pku.edu.cn}
\affiliation{Kavli Institute for Astronomy and Astrophysics, Peking University, Beijing 100871, People's Republic of China}
\author{Zi-Qi He}
\affiliation{Department of Astronomy, School of Physics, Peking University, Beijing 100871, People's Republic of China; E-mail: jjjin@pku.edu.cn; wuxb@pku.edu.cn}
\affiliation{Kavli Institute for Astronomy and Astrophysics, Peking University, Beijing 100871, People's Republic of China}
\author{Qin-Chun Ma}
\affiliation{Department of Astronomy, School of Physics, Peking University, Beijing 100871, People's Republic of China; E-mail: jjjin@pku.edu.cn; wuxb@pku.edu.cn}
\affiliation{Kavli Institute for Astronomy and Astrophysics, Peking University, Beijing 100871, People's Republic of China}
\author{Yu-Xuan Pang}
\affiliation{Department of Astronomy, School of Physics, Peking University, Beijing 100871, People's Republic of China; E-mail: jjjin@pku.edu.cn; wuxb@pku.edu.cn}
\affiliation{Kavli Institute for Astronomy and Astrophysics, Peking University, Beijing 100871, People's Republic of China}
\author{Rui Zhu}
\affiliation{Department of Astronomy, School of Physics, Peking University, Beijing 100871, People's Republic of China; E-mail: jjjin@pku.edu.cn; wuxb@pku.edu.cn}
\affiliation{Kavli Institute for Astronomy and Astrophysics, Peking University, Beijing 100871, People's Republic of China}
\author{Yan-xia Zhang}
\affiliation{National Astronomical Observatories, Chinese Academy of Sciences, Beijing 100012, People’s Republic of China}
\author{Hai-long Yuan}
\affiliation{National Astronomical Observatories, Chinese Academy of Sciences, Beijing 100012, People’s Republic of China}
\author{Zhi-ying Huo}
\affiliation{National Astronomical Observatories, Chinese Academy of Sciences, Beijing 100012, People’s Republic of China}

\begin{abstract}
	
We report the fourth installment in the series of the Large Sky Area Multi-Object Fiber Spectroscopic Telescope (LAMOST) Quasar Survey, which includes quasars observed between September, 2017 and June, 2021. There are in total 13,066 quasars reliably identified,  of which 6,685 are newly discovered that are not reported in the SDSS DR14 quasar catalog or Million Quasars catalog.
Because LAMOST does not provide accurate absolute flux calibration, we re-calibrate the spectra with the SDSS/Pan-STARRS1 multi-band photometric data. The emission line properties of H$\alpha$, H$\beta$, Mg\,{\sc ii} and C\,{\sc iv}, and the continuum luminosities are measured by fitting the re-calibrated spectra. We also estimate the single-epoch virial black hole masses ($\rm M_{BH}$) using the derived emission line and continuum parameters. This is the first time that the emission line and continuum fluxes were estimated based on LAMOST re-calibrated quasar spectra. The catalog and spectra for these quasars are  available online. After the nine-year LAMOST quasar survey, there are in total 56,175 identified quasars, of which 24,127 are newly discovered. The LAMOST quasar survey not only discovers a great number of new quasars, but also provides a database for investigating the spectral variability of the quasars observed by both LAMOST and SDSS, and finding rare quasars including changing-look quasars and broad absorption line quasars.

\end{abstract}

\keywords{catalogs - quasars: emission lines - quasars: general - surveys}

\section{Introduction} \label{sec:intro}

Quasars are a class of active galactic nuclei (AGNs), which are powered by accretion onto the supermassive black holes (SMBHs). Quasars are the most luminous and energetic celestial objects in the universe that can emit radiations over a broad range of wavelength from radio to $\gamma-$ray \citep{1993ARA&A..31..473A}.  Quasars have long been used in a variety of astrophysical studies, such as revealing the growth of SMBHs across cosmic time and the evolution connections to their host galaxies \citep{1998AJ....115.2285M,2000ApJ...539L...9F,2000ApJ...539L..13G,2001AIPC..586..363K,2001MNRAS.320L..30M,2002ApJ...574..740T,2021ApJ...918...22L,2021ApJ...922..142L}, probing the distribution of interstellar and intergalactic medium (ISM and IGM), and tracing the large-scale structure of the early universe \citep{2007ApJ...655..735H,2001AJ....122.2850B}. In addition, quasars are primary celestial references because they are distant extragalactic sources with extremely small proper motions \citep{1998A&A...331L..33F,2009A&A...505..385A}.

Since the first discovery of the quasars in 1963 \citep{1963Natur.197.1040S}, huge efforts have been undertaken to find more quasars. Quasars  can be separated from normal galaxies and stars due to their unique features, such as the characteristic spectral energy distribution, high luminosities, variability properties and radiations at multi-wavelengths. 

The most common method to select quasar candidates is based on the multi-color properties. Particularly, quasars at $ z < 2.2$ have strong UV and optical emissions that distinguish them from normal stars in the color-color and color-magnitude diagrams \citep{2000AJ....120.1167F, 2009ApJS..180...67R,2007AJ....134..102S}. For example, two of the most productive quasar surveys, the Sloan Digital Sky Survey (SDSS; \citealt{2011ApJS..194...45S,2012A&A...548A..66P,2018A&A...613A..51P}) and the Two-Degree Fields (2dF) Quasars Redshift Survey \citep{2000MNRAS.317.1014B} used the optical photometric data to select quasar candidates. However, such optical color selection methods are systematically incomplete at $2.2 < z< 3.0$, especially at $ z = 2.7$ as the quasars in this redshift range have similar colors to those of stellar objects \citep{1999AJ....117.2528F,2002AJ....123.2945R,2006AJ....131.2766R,2007AJ....134..102S}. An efficient way of identifying missing quasars at $2.2 < z< 3.0$ is using the K-band photometry from the UK Infrared Telescope (UKIRT) Infrared Deep Sky Survey (UKIDSS; \citealt{2007MNRAS.379.1599L}), because quasars at $ 2.2 < z< 3.0$ have an excess in the near-infrared K-band when compared to stellar objects \citep{2000MNRAS.312..827W,2002MNRAS.337.1153S,2008MNRAS.386.1605M,2008MNRAS.389..407S}. Thanks to the larger survey area, the all sky survey \textit{Wide-field Infrared Survey Explorer} (\textit{WISE}; \citealt{2010AJ....140.1868W,2012yCat.2311....0C}) shines more light on quasar surveys. It has been demonstrated  that the SDSS/UKIDSS and SDSS/WISE colors can significantly improve the efficiency of quasar selection \citep{2010MNRAS.406.1583W,2012AJ....144...49W}. Other main quasar candidate selection techniques based on the physical characteristics of quasars include: multi-wavelength (X-ray/radio) data matching \citep{1963Natur.197.1040S,2002ApJ...569L...1S,2004MNRAS.353..211C,2011ApJ...736...57Z,2016AJ....151...24A}; variability-based selection \citep{2012ApJ...753..106M,2016AJ....151...24A,2016A&A...587A..41P}; slitless-spectroscopy survey for broad emission line features \citep{2008A&A...487..539W,1986MitAG..67..174C} and proper motion \citep{2020A&A...644A..17H}.

Recently, various data-mining algorithms based on photometric data have also been applied to select quasar candidates, including the Kernel Density Estimation (KDE; \citealt{2004ApJS..155..257R,2009ApJS..180...67R}), the extreme deconvolution method (XDQSO; \citealt{2011ApJ...729..141B}), support vector machine (SVM; \citealt{2012MNRAS.425.2599P}), Gaussian mixtrue model \citep{2019MNRAS.490.5615B}, boosting algorithm (e.g. XGBoost; \citealt{2019MNRAS.485.4539J}) and deep learning \citep{2010A&A...523A..14Y,2018A&A...611A..97P}. For example, the KDE method has been exploited in the SDSS-III Baryon Oscillation  Spectroscopic survey (BOSS; \citealt{2012ApJS..199....3R}), and transfer learning is adopted for finding quasars behind the Galactic plane (GPQs; \citealt{2021ApJS..254....6F}). 

Quasar candidates selected from various methods still need to be spectroscopically identified. This paper presents the results of the Large Sky Area Multi-object Fiber Spectroscopic Telescope (LAMOST) Quasar Survey conducted between September, 2017 and June, 2021. This is the fourth installment in the series of LAMOST quasar survey, after data release 1 (DR1; \citealt{2016AJ....151...24A}, hereafter Paper I), data release 2 and 3 (DR2\&3; \citealt{2018AJ....155..189D}, hereafter Paper II) and data release 4 and 5 (DR4\&5, \citealt{2019ApJS..240....6Y}, hereafter Paper III). In this paper, the candidate selection, spectroscopic survey and quasar identification are briefly reviewed in Section 2. Spectral measurements and $\rm M_{BH}$ estimations for identified quasars are described in Section 3. The description of the quasar catalog and parameters released are presented in Section 4. At last, the summary and discussion are given in Section 5. We adopt the cosmology parameter $\rm H_{0}=70$ km $\rm s^{-1}$ $\rm Mpc^{-1}$ and a flat universe with $\rm \Omega_{M}=0.3$ and $\rm \Omega_{\Lambda}=0.7$.

\section{Survey Outline} \label{sec:survey_outline}

LAMOST, also known as the Guoshoujing Telescope, is a quasi-meridian reflection Schmidt telescope with an effective aperture varies from 3.6 to 4.9 meters \citep{1996ApOpt..35.5155W,2004ChJAA...4....1S,2012RAA....12.1197C,2012RAA....12..723Z}. It is located at Xinglong Observatory, China and
 has a $\rm 5^{\circ}$ (diameter) field of view. LAMOST is equipped with 4000 fibers with $\rm 3.3^{\prime \prime}$ diameter, which are mounted on the focal plane and connected to 16 spectrographs. Each spectrum is divided into a blue channel (3700 $\sim$ 5900 $\rm \AA$) and a red channel (5700 $\sim$ 9000 $\rm \AA$), with an overlapping region between the two channels at 5700-5900 $\rm \AA$. The spectra were observed under the low-resolution mode with a spectral resolution of R $\sim$ 1000-2000 over the entire wavelength range.
 
 After the commissioning from 2009 to 2010, LAMOST began a pilot survey in 2011 \citep{2012RAA....12.1243L}. The LAMOST regular survey starts from September, 2012, which has two major parts \citep{2012RAA....12..723Z}: the LAMOST Experiment for Galactic ExtraGAlactic Survey (LEGAS), and the LAMOST Experiment for Galactic Understanding and Exploration survey (LEGUE).  
 The LAMOST quasar survey was conducted under the LEGAS. 
 The exposure time is 
 adjusted according to the apparent magnitude of targets and observation conditions. The typical value of the total exposure time for a target is $\sim$ 90 minutes, which is equally divided into three sub-exposures. Although the LAMOST quasar survey used only a small fraction of the available observing time due to the limitations of LAMOST site (e.g., weather conditions, poor seeing and bright sky background), LAMOST has still collected useful data and identified more than 40,000 quasars, about half of which are new discoveries, during the first five years. 
 
\subsection{Target Selection}

The methods used to select the quasar candidates for LAMOST quasar survey are described in detail in \cite{2010MNRAS.406.1583W}, \cite{2012AJ....144...49W}, \cite{2012MNRAS.425.2599P}, Paper I, II and III. Here, we just give a brief description of  the candidate selection.

The primary selection for quasar candidates is based on the photometric data of SDSS \citep{2012ApJS..203...21A}, and the magnitudes we used here are the SDSS point-spread function (PSF) magnitudes with the Galactic extinction corrected \citep{1998ApJ...500..525S}.  First, only point sources are selected to exclude galaxies. We notice that we may miss some quasars at low redshifts with extended morphology due to this limitation. Second, the targets should be brighter than i=20 to avoid too low signal-to-noise ratio (S/N), and fainter than i=16 to avoid saturation and contamination with neighbor fibers. Various methods are then applied to further  separate quasar candidates from stars. Most of the quasar candidates are selected based on the optical-infrared colors (SDSS-UKIDSS/WISE), as has been described in \cite{2010MNRAS.406.1583W} and \cite{2012AJ....144...49W}.  A few data-mining algorithms are also used to selected quasar candidates, such as  SVM classifiers \citep{2012MNRAS.425.2599P}, extreme deconvolution method (XDQSO; \citealt{2011ApJ...729..141B}), and KDE \citep{2009ApJS..180...67R}. In addition, some quasar candidates are selected by cross-matching SDSS photometry with the detected sources in X-ray surveys (XMM-Newton, Chandra, ROSAT) and radio surveys (FIRST, NVSS).

Although some of the selected candidates have already been identified by SDSS after our target selections, we include them in the LAMOST survey, which will be helpful to investigate the spectroscopic variability of quasars and find unusual quasars.

\subsection{Pipeline for Data Reduction}
The raw CCD images obtained from observations were reduced by the LAMOST two-dimensional (2D) pipeline and one-dimensional (1D) pipeline, which are described in \cite{2015RAA....15.1095L}. The 2D pipeline is applied to extract 1D spectra from the raw data through a sequence of procedures, including dark and bias subtraction, flat-field correction, cosmic-ray removal, spectral tracing and extraction,  sky subtraction, wavelength calibration, merging sub-exposure, relative flux-calibrate and combining blue and red spectra \citep{2012RAA....12.1243L}. Then through the 1D pipeline, these 1D spectra are automatically classified into four primary categories: ``STAR", ``GALAXY", ``QSO", and ``Unknown" by template matching. The final spectra are available at the LAMOST Data Archive Server \footnote{http://www.lamost.org/lmusers/} (DAS).  

The 1D pipeline classification is not trustworthy for the ``Unknown" type.  The main reason is that these ``Unknown" spectra are taken under non-photometric conditions, e.g., varying seeing and/or cloudy weather. In addition, unstable efficiencies  of some fibers also contribute to the high fraction of ``Unknown" objects. 
In the LAMOST early data release, only $\rm \sim 14\%$ of the observed quasar candidates are classified as QSO, STAR, or GALAXY by the pipeline, while the majority of the spectra are categorized as ``Unknown"  (Paper I). Such high fraction of unrecognizable objects is due to the poor spectral quality in the early data release. 
Both the varying seeing and the non-classical dome of the telescope significantly affect the spectral quality, and the targets with magnitude fainter than $\rm i=20$ are challenging for the LAMOST LEGAS survey.
Fortunately, the candidate selection in the regular survey was improved when compared to those in the pilot survey, and the LAMOST data reduction pipeline has been updated for a better performance of spectral classification. As a result, the fraction of candidates classified as ``QSO" keeps relatively high in later regular survey ($55.9\%$ in Paper II, $62.3\%$ in Paper III and $77.0\%$ in this work).

\subsection{Quasar Identification}

In this work, the quasars are identified by visual inspections. In addition to the observed spectra of quasar candidates, the spectra that are classified as ``QSO" by the 1D pipeline but not included in the input quasar candidate catalog also need visual inspections. With the help of a Java program ASERA \citep{2013A&C.....3...65Y}, we visually inspect these spectra based on the typical quasar emission lines. Each spectrum is inspected by at least two persons to check if the spectral features match the quasar template. The objects that are misclassified by the 1D pipeline are rejected or reclassified. The redshift of each identified quasar is determined when one or more available typical quasar emission lines (e.g., as H$\alpha$, H$\beta$, O{\,\sc iii}$\lambda5007$, Mg\,{\sc ii}, C\,{\sc iii} and C\,{\sc iv}) are best matched with the templates.
The ``ZWARNING = 1" flag indicates there is only one emission line available. The quasars that overlap with M31/M33 and Galactic-anti-center extension region (GACext) will be published elsewhere (see \citealt{2010RAA....10..612H,2013AJ....145..159H,2015RAA....15.1438H}), and are not included in our final quasar catalog. Finally, there are in total 13,066 visually confirmed quasars from data release 6 to 9 in the quasar catalog. 9827 of them are not included in LAMOST quasar survey candidate catalog (updated in 2017, and a new version including PS1 quasar candidates is still in preparing) but identified as quasars. Among the 13066 identified quasars, after excluding known ones in common with SDSS DR14 quasar catalog or Million Quasars catalog (Milliquas v7.5\footnote{http://www.quasars.org/milliquas.htm}; \citealt{2021arXiv210512985F}), the remaining 6,685 are newly discovered. Since the LAMOST DR6 quasar survey was finished in the same year as the SDSS DR14 quasar catalog was published, the 417 quasars in LAMOST DR6 that are in common with SDSS DR14 are considered as independently discovered by LAMOST. Therefore, there are in total 7102 quasars that were independently discovered by LAMOST.  
 The result of quasar identification is summarized in Table~\ref{table:catalog}. The SDSS DR16 quasar catalog \citep{2020ApJS..250....8L} was published in September, 2020 and the observations of most objects in our survey were completed before that. Therefore we no longer make a comparison between SDSS DR16Q and this work.  We caution, however, the 1257 of newly discovered LAMOST quasars were reported in SDSS DR16Q. 
With a large number of repeat spectral observations of SDSS and LAMOST, we can investigate the spectroscopic variability of quasars on both short and long time scales. Moreover, these multi-epoch spectra give us a good chance to search for unusual AGNs such as changing-look AGNs (CL-AGNs; e.g. \citealt{2016ApJ...821...33R,2019ApJ...874....8M,2018ApJ...862..109Y,2019ApJ...887...15W,2019ApJ...883L..44G}) and uncover the possibl physical mechanism behind them \citep{2019ApJ...874....8M,2019ApJ...883...31F,2022ApJ...926..184J}. 


\begin{table}[htbp]
	\centering
	\caption{The result of LAMOST quasars survey in DR6, 7, 8 and 9.}
	\begin{tabular}{lccccc}
		\hline 
		\hline 
		&DR6&DR7&DR8&DR9&Total\\
		\hline 
		Total&4275&2294&3883&2614&13066\\
		Independent&2245&879&2223&1755&7102\\
		New&1828&879&2223&1755&6685\\
		\hline 
	\end{tabular}
	\label{table:catalog}
\end{table}

\begin{figure}[!htb]
	\begin{center}	
		\includegraphics[page=1,scale=0.39]{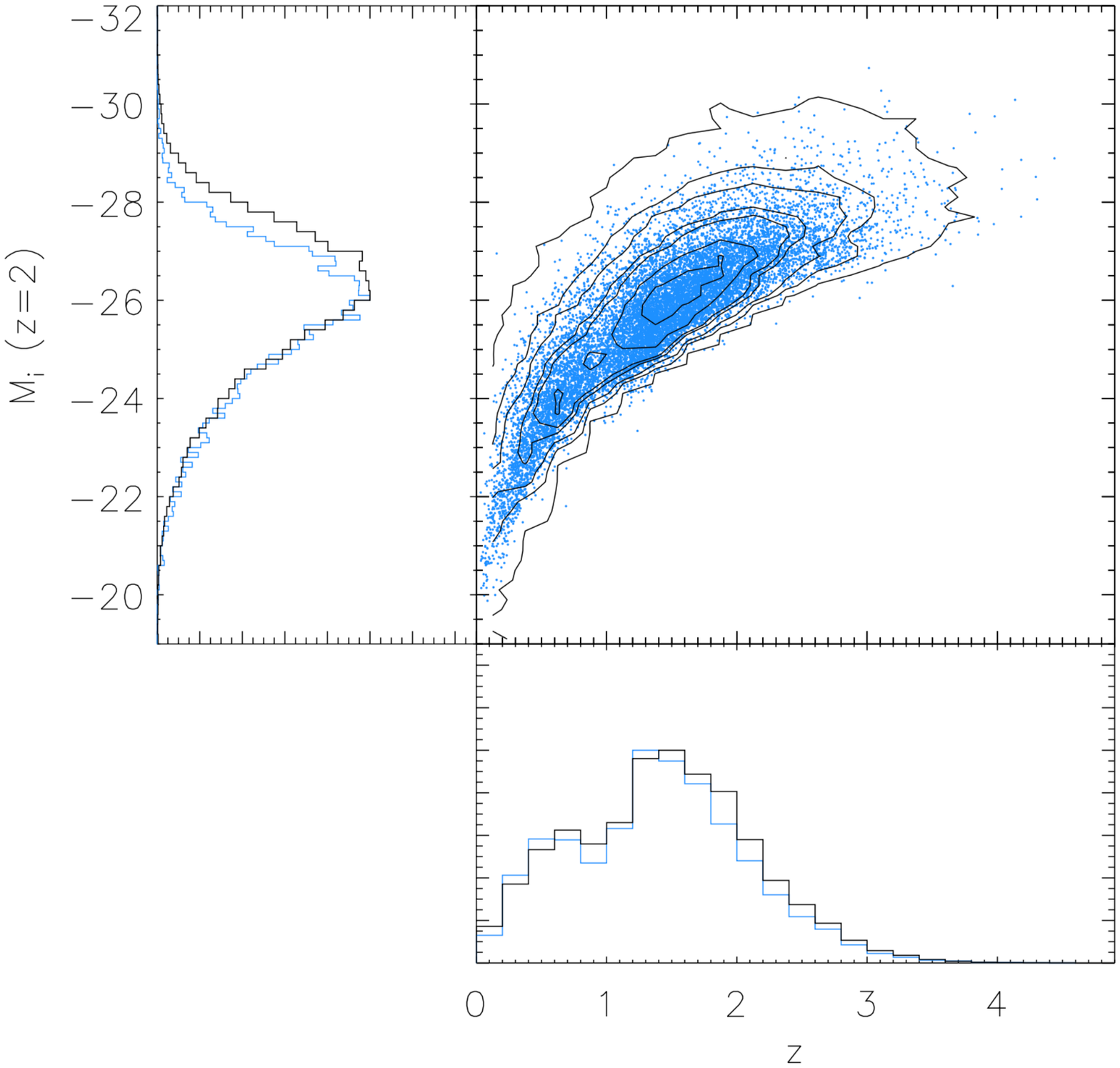}   
		\caption{The distribution in the magnitude-redshift space for the visually confirmed quasars for previous LAMOST quasar survey (black contours) and in DR6-9 (blue). The absolute magnitudes $\rm M_{i} (z=2)$ are normalized at z=2, following the K-correction of \cite{2006AJ....131.2766R}. The left and bottom panels show the absolute magnitude and redshift distributions, respectively.}
	\end{center}
	\label{fig:z_MI} 
\end{figure}

\begin{figure}[!htb]
	\begin{center}	
		\includegraphics[page=1,scale=0.49]{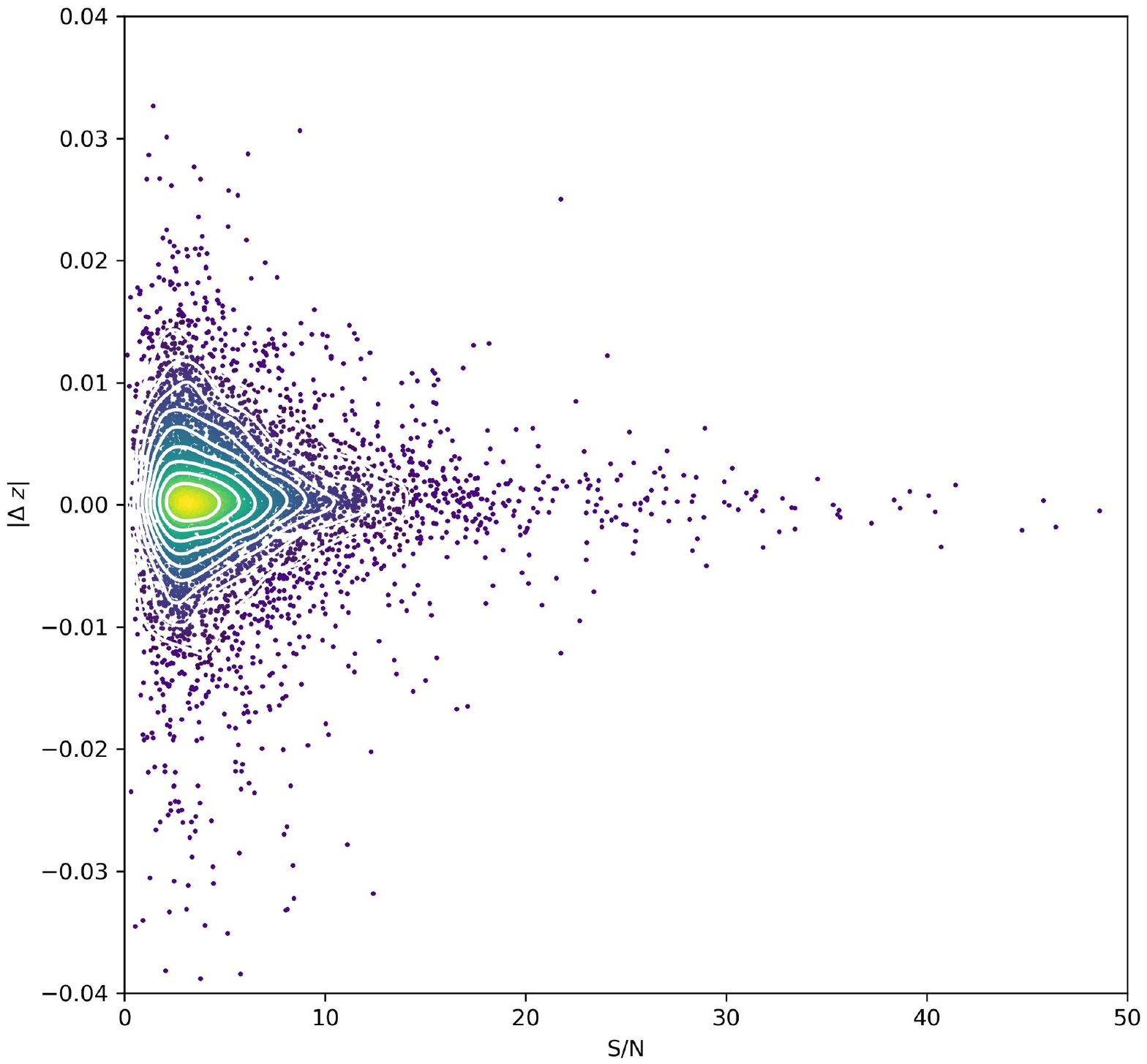}   
		\caption{The distribution of redshift difference ($\Delta z$) for common quasars between this work and SDSS versus LAMOST spectral S/N.}
	\end{center}
	\label{fig:snr_detlaZ} 
\end{figure}

Figure ~\ref{fig:z_MI} shows the distribution of the redshift and absolute luminosity, which is represented by the K-corrected i-band absolute magnitude $\rm M_{i}$ $\rm(z=2)$, normalized at z = 2 \citep{2006AJ....131.2766R}. As can be seen, there is a drop in the redshift distribution at $\rm z \sim 1$, which is similar to the previous results (Papers I, II and III). This drop is the result of inefficient identification in this redshift range when the emission line Mg\,{\sc ii} moves into the overlapping region of the blue and red channels of the spectrograph. For the sources observed by both SDSS and LAMOST, only 88 of them have redshift difference ($\rm \Delta z = z_{LAMOST} - z_{SDSS}$) greater than 0.1. The difference mainly comes from the misidentification of emission lines in LAMOST spectra due to the low S/N. As shown in Figure ~\ref{fig:snr_detlaZ}, it is clear that as the S/N decreases, the $\Delta z$ increases. Another reason for the redshift difference is that we estimated the redshift based on the strongest typical emission line, while the redshift values in SDSS are measured with a few difference approaches, such as principal component analysis (PCA) or Mg\,{\sc ii} emission line \citep{2018A&A...613A..51P}.

\begin{figure*}[!htb]
	\begin{center}	
		\includegraphics[page=1,scale=0.6]{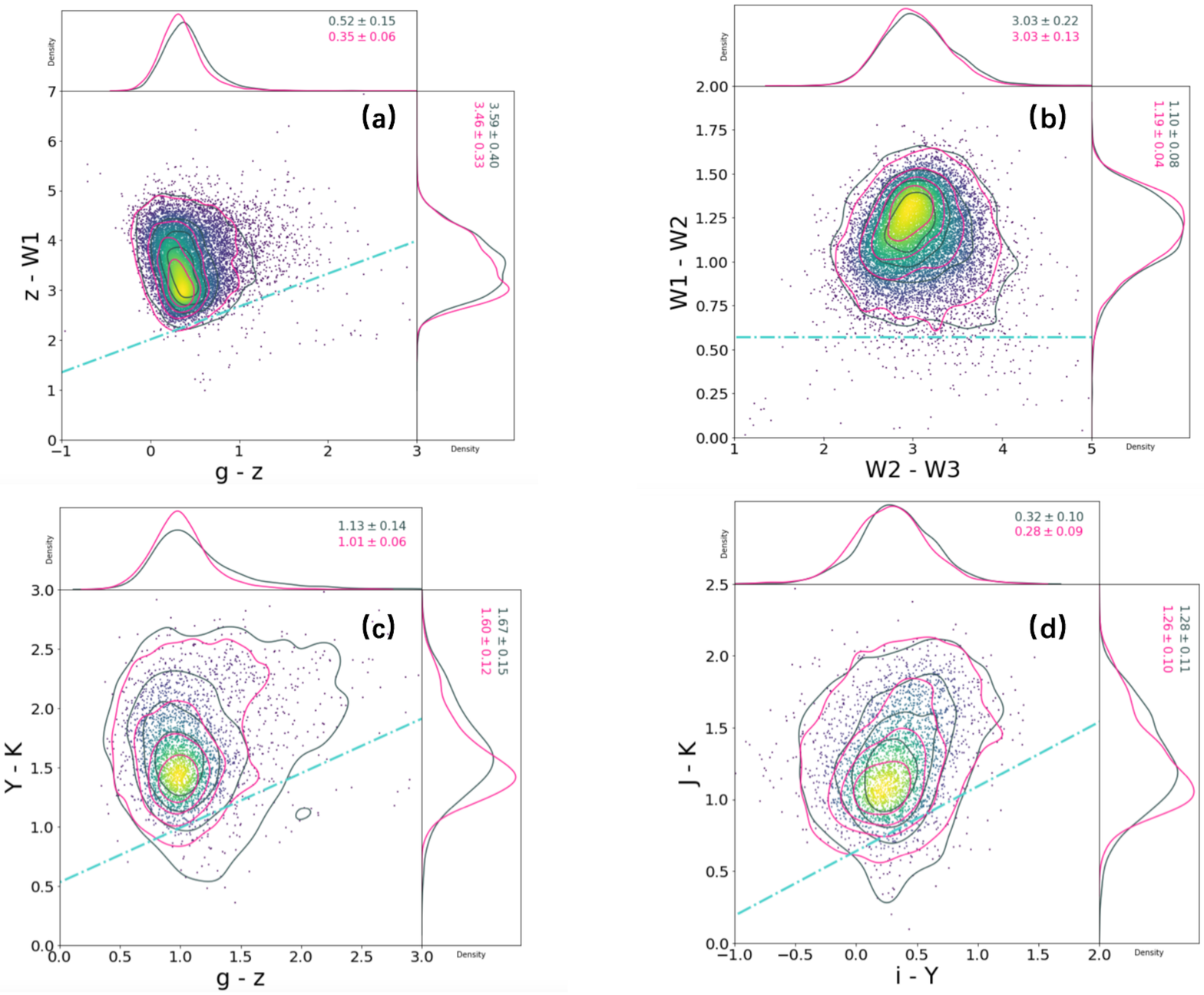}   
		\caption{The distributions of identified LAMOST quasars in the SDSS-WISE/UKIDSS color diagram. The dash-dotted lines indicate the criteria used in the LAMOST QSO survey \citep{2010MNRAS.406.1583W,2012AJ....144...49W}.  The WISE and UKIDSS magnitudes are in Vega magnitudes. The SDSS magnitudes in the panel (a) and (b) are plotted in AB magnitudes and those in the panel (c) and (d) are plotted in Vega magnitudes. The SDSS AB magnitudes can be converted to Vega magnitudes using the following scaling \citep{2006MNRAS.367..454H}: $\rm g = g(AB)+0.103$, $\rm r=r(AB)-0.146$, $\rm i=i(AB)-0.366$, $\rm z=z(AB)-0.533$. The contours in pink show the distribution for common quasars between this work and SDSS, while the contours in gray show the distribution for unique quasars identified in this work. The mean ($\mu$) and dispersion ($\sigma$) of each distribution are tabulated in corresponding plots.}
	\end{center}
	\label{fig:color} 
\end{figure*}

We present the SDSS-WISE/UKIDSS color-color distributions for these identified quasars in Figure ~\ref{fig:color}. It is clear that the most ($\rm 99\%$) of the identified quasars locate in the selection regions obtained with the optical-infrared color, suggesting that the optical-infrared color selection is a very promising method for selecting quasars. Generally, the quasars uniquely identified by LAMOST are slightly redder in colors than those commonly identified by LAMOST and SDSS. The reasons for these color differences may be that the SDSS quasars are mainly selected by optical colors or optical variability \citep{2002AJ....123.2945R,2015ApJ...806..244M,2015ApJS..221...27M,2016A&A...587A..41P}.

\section{Spectral Analysis}

In this section, we describe the spectral analysis, which includes the absolute flux-calibration, the measurements of typical quasar emission lines, and the estimations of $\rm M_{BH}$.  

\subsection{Absolute Flux Calibration}
We note that the LAMOST is designed as a spectroscopic survey without photometric measurements for the observed targets, and there may be not enough standard stars for a given LAMOST field especially at high Galactic latitude. Thus this instrument can only provide relative flux calibration rather than absolute flux calibration  \citep{2012RAA....12..453S,2015MNRAS.448...90X}.  As mentioned above, the process of relative flux calibration is the final step of  2D pipeline. In the first step of relative flux-calibration, the A and F type stars with high quality spectra are selected as  pseudo-standard stars for each spectrograph, and are used to calibrate both the blue and red spectrograph arms. The effective temperatures of these stars are estimated using the Lick spectral index grid \citep{1972PASP...84..161R,1998ApJS..116....1T}. Then the spectral response curve (SRC) of each spectrograph is obtained by dividing the observed continuum from the data by the physical pseudo-continuum for the star. Further, these SRCs are applied to all other fiber spectra to calibrate them. Finally, the red and blue spectra are combined by stacking the points with corresponding wavelength using B-spline function with inverse-variance \citep{2012RAA....12..453S,2015RAA....15.1095L}.

Only with the relatively flux-calibrated spectra, we can not estimate the emission line flux, as well as the continuum luminosity for LAMOST quasars. However, the absolute calibration can be achieved by scaling the relative flux calibrated spectra to the photometric measurements \citep{2015MNRAS.448...90X}.
In this work, we try to achieve the absolute calibration by scaling each spectrum to the corresponding broad-band photometric measurements.  The broad-band photometry used in this work are the PSF magnitudes from the SDSS \citep{2000AJ....120.1579Y} or Pan-STARRS1 \citep{2016arXiv161205560C,2020ApJS..251....7F}. First, we cross-match the LAMOST quasars with the SDSS photometric database with a 3$^{\prime \prime}$ matching radius. The sources outside the SDSS footprint are then cross-matched with Pan-STARRS1 with the same matching radius\footnote{The spectra of eight quasars that do not have reliable SDSS or Pan-STARRS1 photometric information are not flux-calibrated in this work.}. Due to the limitation of spectral wavelength coverage, we only use g,r,i-band during the calibration. The magnitudes in these three bands are converted into the flux density $f_{\lambda}$ at the effective wavelength of each filter. Next, we fit each quasar spectrum with the flux densities in the three bands. 

\begin{figure*}[!htb]
	\begin{center}	
		\includegraphics[page=1,scale=0.45]{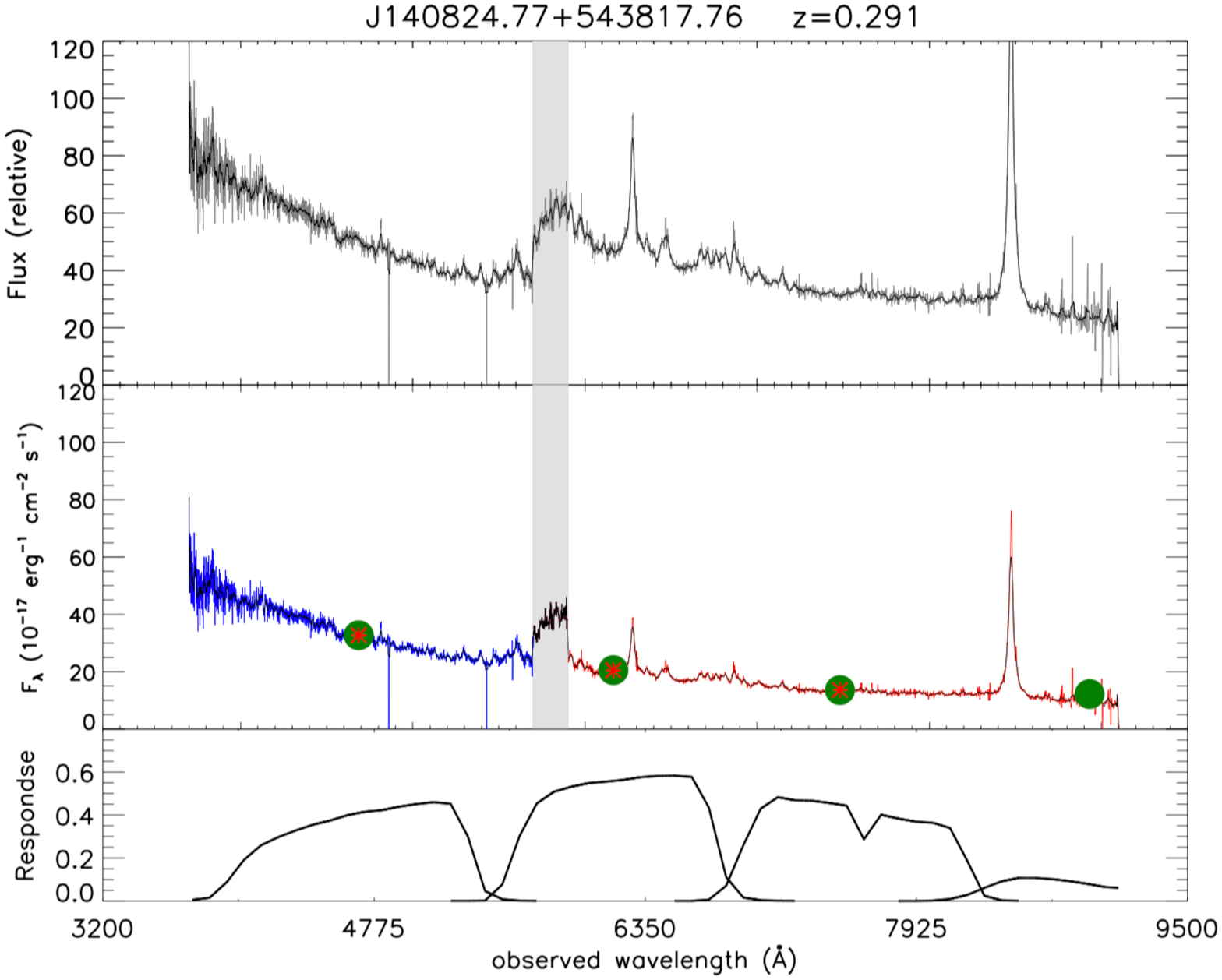}   
		\caption{An example of absolute flux-calibration for the blue- and red-arm LAMOST spectra. Top panel: The original spectrum only with the relative flux calibration. The gray area represents the blue-red overlapping region which is masked during the fitting. A connection defect can be observed at this region. Middle panel: The spectrum after the absolute flux-calibration. The green dots represent the flux densities in the g, r, i, z-bands, and the asterisks mark the flux densities that are used during the fitting. The lines in blue and red represent piecewise fits to the asterisks. The connection defect is improved after re-calibration. The bottom panel shows the filter curves for the SDSS in g, r, i, z-band. It is clear that the z-band is not fully covered by the spectrum, so the photometric data in z-band is not used during the spectral fitting.}
	\end{center}
	\label{fig:RedBlue} 
\end{figure*}

Since the spectra from the blue and red channels are relative flux-calibrated separately in LAMOST 2D pipeline, the absolute flux-calibration is also applied to the blue and red channels separately. The  fits are based on the IDL routines in the {\tt MPFIT}  package \citep{2009ASPC..411..251M}, which performs the $\chi^{2}$ minimization using the Levenberg-Marquardt method. Examples of the fitting results are presented in Figure ~\ref{fig:RedBlue}. The released LAMOST spectrum (the gray spectrum in the top panel) only has the relative flux distribution without unit. After the absolute flux-calibration, the flux density of LAMOST spectrum has the unit of $\rm erg$ $\rm cm^{-2}$ $\rm s^{-1}$ $\rm \AA^{-1}$.  Actually, there is only a scale applied to the LAMOST released spectra separately in the blue and red channels corresponding to the g, r, i bands, so the spectral shapes in the blue and red channels  are not changed during the process.

As we mentioned before, one step in LAMOST 2D pipeline is connecting the spectra in blue and red channel to each other. However, in some cases, this procedure produces strange shapes in the continuum with connection defect, which is shown clearly in the top panel of Figure ~\ref{fig:RedBlue} as an example. This defect may cause unpredictable errors in the subsequent spectral fitting process. Fortunately, the strange shapes caused by the defect can be improved by the process of absolute flux-calibration as the blue and red channels are re-connected corresponding to the photometry data. As shown in the  middle panel of Figure ~\ref{fig:RedBlue}, it is clear that the shape of continuum in the blue and red channels conform the common power-law shape after the absolute flux calibration.

Quasars usually show optical variabilities of 0.1-0.2 mag, which introduce additional  uncertainties to the absolute flux-calibration. However, the spectra without the absolute flux calibration information can not be used to obtain important quantities including continuum luminosity, $\rm M_{BH}$ and emission line flux.  In the previous paper of LAMOST quasar survey (Papers I, II and II), the spectra are not absolute-flux-calibrated, the continuum luminosity is inferred from the model fitting with the SDSS photometric data, and there is no emission line flux information in the published catalogs. Additionally,  a small fraction of spectra ($\sim$ 17\% in Paper II and $\sim$ 6\% in Paper III) cannot be fitted properly due to the connection defect near the overlapping region, which is solved by the absolute flux-calibration in this work. Despite the uncertainties of our absolute flux-calibration, it nevertheless helps understand more about the central BHs of these quasars.

\subsection{Spectral fitting}
Here we describe the fitting procedures for LAMOST quasar spectra. Before the fitting, each absolute flux calibrated spectrum is corrected for the Galactic extinction using the reddening map \citep{1998ApJ...500..525S} and the Milky Way extinction law of \cite{1999PASP..111...63F} with $\rm R_{V} = 3.1$, and then transformed into the rest-frame using the redshift obtained from the visually inspected.

Then the spectra are fitted by the publicly avaiable multicomponent spectral fitting code {\tt pyQSOFit} \citep{2018ascl.soft09008G} and a wrapper package based on it ({\tt QSOFITMORE}; \citealt{zenodo.5810042}). The {\tt pyQSOFit} performs the $\rm \chi^2$ fits, using the estimated errors per pixel that come from the reduction pipeline. A detailed description of the code and its application can be found in  \cite{2018ascl.soft09008G}, \cite{2019ApJS..241...34S} and \cite{zenodo.5810042}. 
\subsubsection{Continuum}
The pseudo continuum is fitted by a broken power law ($f$$\rm_{bpl}$) and a Fe\,{\sc ii} model ($f$$\rm_{Fe\,{\sc ii}}$) in the wavelength windows without quasar emission lines
 and outside the LAMOST spectral overlapping region.
 During the fitting, the turning point of the broken power law is fixed at 4661 $\rm \AA$ at rest-frame, which is similar to the value derived from the mean composite quasar spectra in \cite{2001AJ....122..549V}. Many works \citep{1985ApJ...288...94W,2001AJ....122..549V} show that there is an abrupt slope change near 5000 $\rm \AA$ in the quasar continuum. One possible reason to the steeper slope at long-wavelength is the near-infrared inflection, which is presumably caused by the emission from hot dust \citep{1994ApJS...95....1E}. Another possible reason is the contamination from the host galaxies at low-redshift. A trend of greater contribution from starlight with increasing wavelength is expected because the emission from host galaxies contributes larger fraction at longer wavelength \citep{1998MNRAS.301....1S,2001AJ....122..549V}. Both the external factors and a real change in the quasar continuum cause an abrupt change in its continuum slope \citep{2001AJ....122..549V}. In the spectral fitting process, the iron model $f$$\rm_{Fe~\textsc{ii}}$ is

\begin{equation}
f_{\rm Fe~\textsc{ii}} = b_{0}F{\rm_{Fe~\textsc{ii}}}(\lambda,b_{1},b_{2}),
\end{equation}

\noindent where the parameters $b_{0}$, $b_{1}$, $b_{2}$ are the normalization, the full width at half-maximum (FWHM) of Gaussian profile used to convolve the Fe\,{\sc ii} template, and the wavelength shift applied to the Fe\,{\sc ii} template, respectively. The optical Fe\,{\sc ii} template is based on \cite{1992ApJS...80..109B}. The UV Fe\,{\sc ii} template is a modified template consisting of the templates in the wavelength range of 1000-2000 $\rm \AA$ based on \cite{2001ApJS..134....1V}, 2200-3090 $\rm \AA$ based on  \cite{2007ApJ...662..131S}, and 3090-3500 $\rm \AA$ based on \cite{2006ApJ...650...57T}.
A few spectra have peculiar shapes in the continuum. It may be caused by some uncertainties in the SRC, that results from unstable efficiencies of some fibers and poor relative flux calibrations occasionally. Because it is difficult to find a suitable flux standard star for each spectrograph, especially for our extragalactic targets as they are faint and located at high Galactic latitudes.

\begin{figure*}[!htb]
	\begin{center}	
		\includegraphics[page=1,scale=0.65]{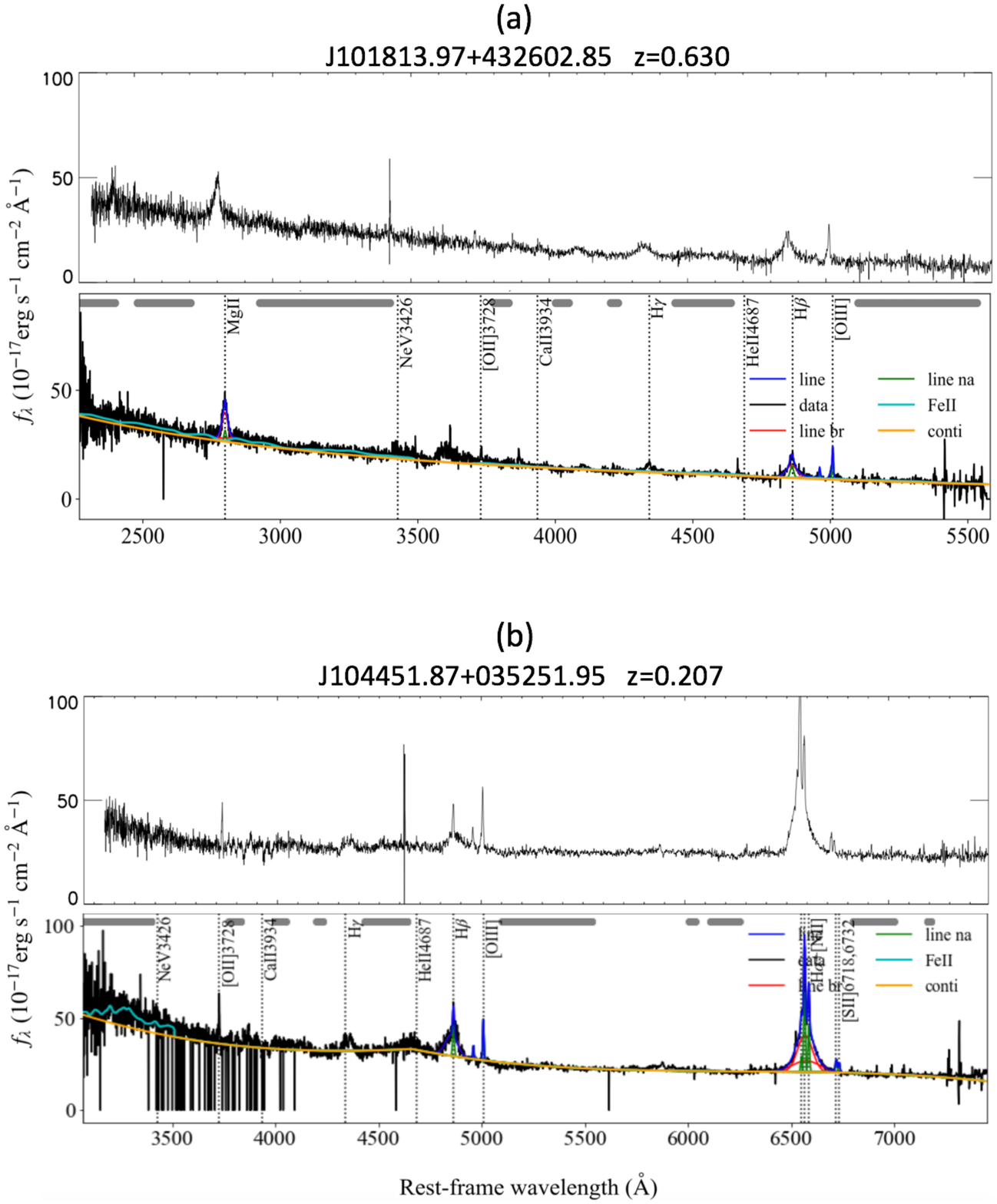}   
		\caption{Two examples for the spectral fitting results with (panel b) and without (panel a) polynomial model. In each figure, the top panel shows the spectrum from SDSS, and the bottom panel shows the spectral fitting results of LMAOST spectrum: the black lines denote the dereddened spectra, yellow lines represent the continuum model of $(f{\rm_{bpl}} + f\rm_{poly})$, and the cyan lines represent  ${\rm {Fe~\textsc{ii}}}$ templates. The details of emission line components will be described in the later part.}
	\end{center}
	\label{fig:poly} 
\end{figure*}

In this case, we add a three-order polynomial model ($f$$\rm_{poly}$) to solve this problem \citep{2020ApJS..249...17R,2022ApJS..261...32F}.  Examples of the fitting results with (and without) polynomial model are presented in Figure ~\ref{fig:poly}. When compared with SDSS spectra, it is clear that there is a peculiar shape in the continuum of LAMOST spectra (as the example at the bottom panel). Only a small fraction of objects require an additional polynomial component ($\lesssim 0.3 \%$).
At last, the pseudo continuum is fitted by two (or three) components:

\begin{equation}
f_{\rm cont} = f{\rm_{bpl}} + f{\rm_{Fe~\textsc{ii}}} + (f\rm_{poly}).
\end{equation}

The host galaxy contamination is negligible for high z ($z \gtrsim 0.5$) or high luminosity ($\rm log L_{5100} \gtrsim$ 44.5) quasars. As for the $z \lesssim 0.5$ low-luminosity quasars, the hosts on average can contribute $\sim 15\%$ to the observed emissions and lead to a $\sim 0.06$ dex overestimation of the 5100 $\rm \AA$ continuum luminosity \citep{2011ApJS..194...45S}. However, due to the limitation of the spectral S/N for faint objects in our catalog, the process of host-subtraction may bring larger uncertainties. Therefore, in this work, the decomposition of the host galaxy is not applied to the spectra.

The fitted pseudo-continuum component is subtracted from the spectrum and the remaining emission-line components are fitted with Gaussian profiles. We focus on four typical quasar emission lines: H$\alpha$, H$\beta$, Mg\,{\sc ii} and C\,{\sc iv}. They are the strongest broad emission lines in the available wavelength range, and are commonly used as virial black hole mass estimators. During the fitting, the parameters we mainly focus are FWHM, equivalent width (EW) and flux. The fitting procedures for each line are described as follows.

\subsubsection{H$\alpha$ line}
The pseudo-continuum-subtracted H$\alpha$-[N\,{\sc ii}]-[S\,{\sc ii}] emission lines are fitted in the rest-frame windows [6350,6800] $\rm \AA$ for objects at $z \lesssim 0.37$. The broad component of H$\alpha$ is modeled by two Gaussian profiles, and the narrow components of H$\alpha$, [N\,{\sc ii}]$\lambda\lambda$6548,6584 and [S\,{\sc ii}]$\lambda\lambda$6716,6731 are each modeled by a single Gaussian profile. The upper limit of FWHM for the narrow components is set to be 900 km $\rm s^{-1}$, which is a commonly used FWHM criterion to separate the narrow and broad lines \citep{2009ApJ...707.1334W,2019A&A...625A.123C,2019ApJ...882....4W}. The line widths and velocity offsets of the narrow lines are tied to each other. The relative flux ratio of the [N\,{\sc ii}]$\lambda\lambda$6548,6584 doublet is fixed to 2.96. Examples of the best-fitting results of H$\alpha$ line are given in the panel (a) of Figure ~\ref{fig:Emission}.

\subsubsection{H$\beta$ line}
The pseudo-continuum-subtracted H$\beta$-[O\,{\sc iii}] emission lines are fitted in the rest-frame window [4600,5100] $\rm \AA$ for objects at $z \lesssim 0.8$. Similar to H$\alpha$, the broad component of H$\beta$ is modeled by two Gaussian profiles, and the narrow component of H$\beta$ is modeled by a single Gaussian profile. The upper limit of FWHM for the narrow components is set to be 900 km $\rm s^{-1}$.  In addition to a single narrow component, the [O\,{\sc iii}]$\lambda\lambda$,4959,5007 double lines require blue wing components as has been suggested by previous studies  (e.g. \citealt{2005AJ....130..381B,2004A&A...413.1087C,2007ApJ...667L..33K,2010MNRAS.403.1759Z,2018A&A...615A..13S}). Therefore each of the [O\,{\sc iii}]$\lambda\lambda$,4959,5007 double lines is modeled by two Gaussians, one for a line core and the other for the blue-shifted wing, and neither of them are tied to the H$\beta$ narrow component. The line widths and velocity offsets of the cores and wings are tied to each other. We constrain the relative flux ratio of [O\,{\sc iii}]$\lambda\lambda$,4959,5007 double lines to be the theoretical ratio of 1:3. Examples of the best-fitting results of the H$\beta$ line are given in the panel (b) of Figure ~\ref{fig:Emission}.

\subsubsection{Mg\,{\sc ii} line}
Fittings of the Mg\,{\sc ii} and C\,{\sc iv} emission lines are sometimes affected by the broad and narrow absorption features. In order to reduce the effect of narrow absorption features, we used ``$\rm rej\_abs = True$''  option of the {\tt QSOFITMORE} code when fitting Mg\,{\sc ii} and C\,{\sc iv} emission lines. The code masks out the 3$\sigma$ outliers below the continuum model, which is useful to reduce the impact of absorption features \citep{2011ApJS..194...45S,2019ApJ...874...22S}.

We fit the Mg\,{\sc ii} emission line for objects at 0.36 $ \lesssim z \lesssim$ 2.1 in the rest-frame wavelength range of [2700,2900] $\rm \AA$. The broad component of Mg\,{\sc ii} is modeled by two Gaussian profiles. As for the narrow component, the situation is more complicated. Some AGNs show the Mg\,{\sc ii}$\lambda\lambda$2796,2803 double lines around the peak, and the FWHM of each component is $\lesssim$ 750 km $\rm s^{-1}$ \citep{2011ApJS..194...45S}. However, such cases are rare and most LAMOST spectra do not have adequate S/N and/or spectral resolution to separate these two components. Additionally, the narrow Mg\,{\sc ii} absorption line can lead to mimicking double peaks. Therefore, in this work we fit the Mg\,{\sc ii} narrow component with a single narrow Gaussian with FWHM upper limit of 900 km $\rm s^{-1}$. Examples of the best-fitting results of Mg\,{\sc ii} line are given in the panel (c) of Figure ~\ref{fig:Emission}.

\subsubsection{C\,{\sc iv} line}
\begin{figure*}[!htb]
	\begin{center}	
		\includegraphics[page=1,scale=0.5]{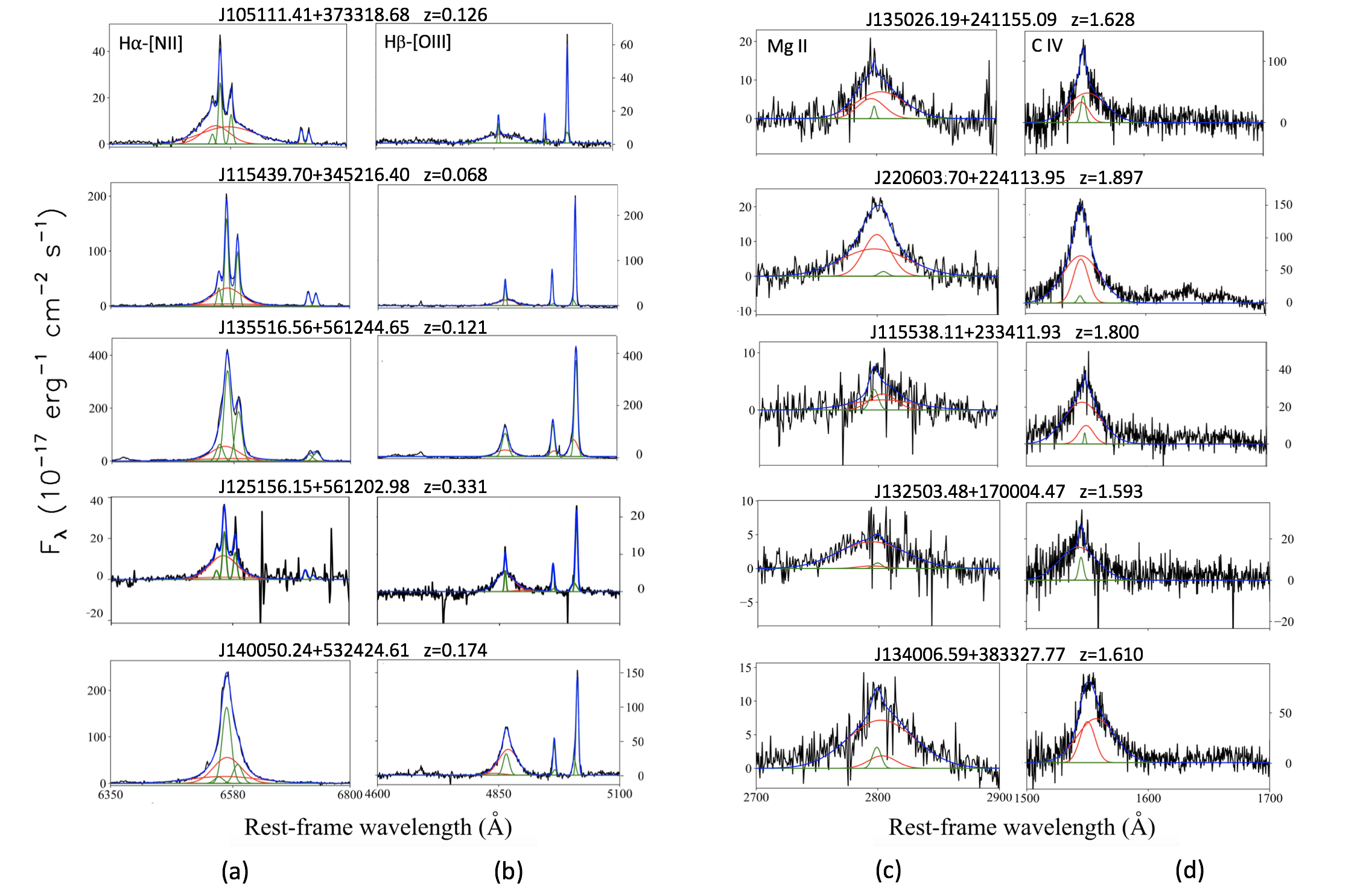}   
		\caption{Examples for the deblending results of H$\alpha$-[N\,{\sc ii}]-[S\,{\sc ii}] (panel a), H$\beta$-[O\,{\sc iii}] (panel b), Mg\,{\sc ii} (panel c) and C\,{\sc iv} (panel d) lines. The black lines represent the extinction-corrected spectra with the continuum subtracted. As for the fitted  emission lines, the broad components are in red while the narrow ones are in green, along with their sum (blue).} 
	\end{center}
	\label{fig:Emission} 
\end{figure*}

We fit the C\,{\sc iv} emission line for objects at 1.5 $ \lesssim z \lesssim$ 4.4 in the spectral rest-frame range of [1500,1700]$\rm \AA$. Similar to other emission lines, the broad component of C\,{\sc iv} line is modeled by two Gaussian profiles. We do not set the upper limit for the FWHM of the narrow component because it is still debatable whether a strong narrow C\,{\sc iv} component exists for most quasars \citep{2011ApJ...742...93A, 2012ApJ...759...44D, 2019ApJS..241...34S}. In addition to the broad and narrow components, the parameters of the entire C\,{\sc iv} profile are also given because: (1) it is not sure whether the narrow component subtraction is feasible for C\,{\sc iv} emission line, and (2) the existing  C\,{\sc iv} virial estimators are calculated with the FWHM from the entire C\,{\sc iv} profiles. Examples of the best-fitting results of C\,{\sc iv} are given in panel (d) of Figure ~\ref{fig:Emission}.

\subsubsection{The Reliability of the Spectral Fitting and Error Estimation}
After the automatic fitting  procedures, we visually inspect the fitting results for each object. The fittings are acceptable for most of the spectra with high S/N. The bad fittings are mainly caused by low S/N of the spectra and the lack of good pixels in the fitting region. A flag is given for each line based on the visual inspection : {\tt LINE\_FLAG = 0} indicates an acceptable fitting and reliable measurement; {\tt LINE\_FLAG = -1} indicates a spurious fitting; {\tt LINE\_FLAG = -9999} indicates that there are not enough good pixels in the fitting region due to the limitation of spectral quality or wavelength region. The broad absorption line (BAL) features can also affect the fitting results. Those BAL features at Mg\,{\sc ii} and/or C\,{\sc iv} are marked with {\tt BAL\_FLAG = 1}. 

\begin{figure}[!htb]
	\begin{center}	
		\includegraphics[page=1,scale=0.39]{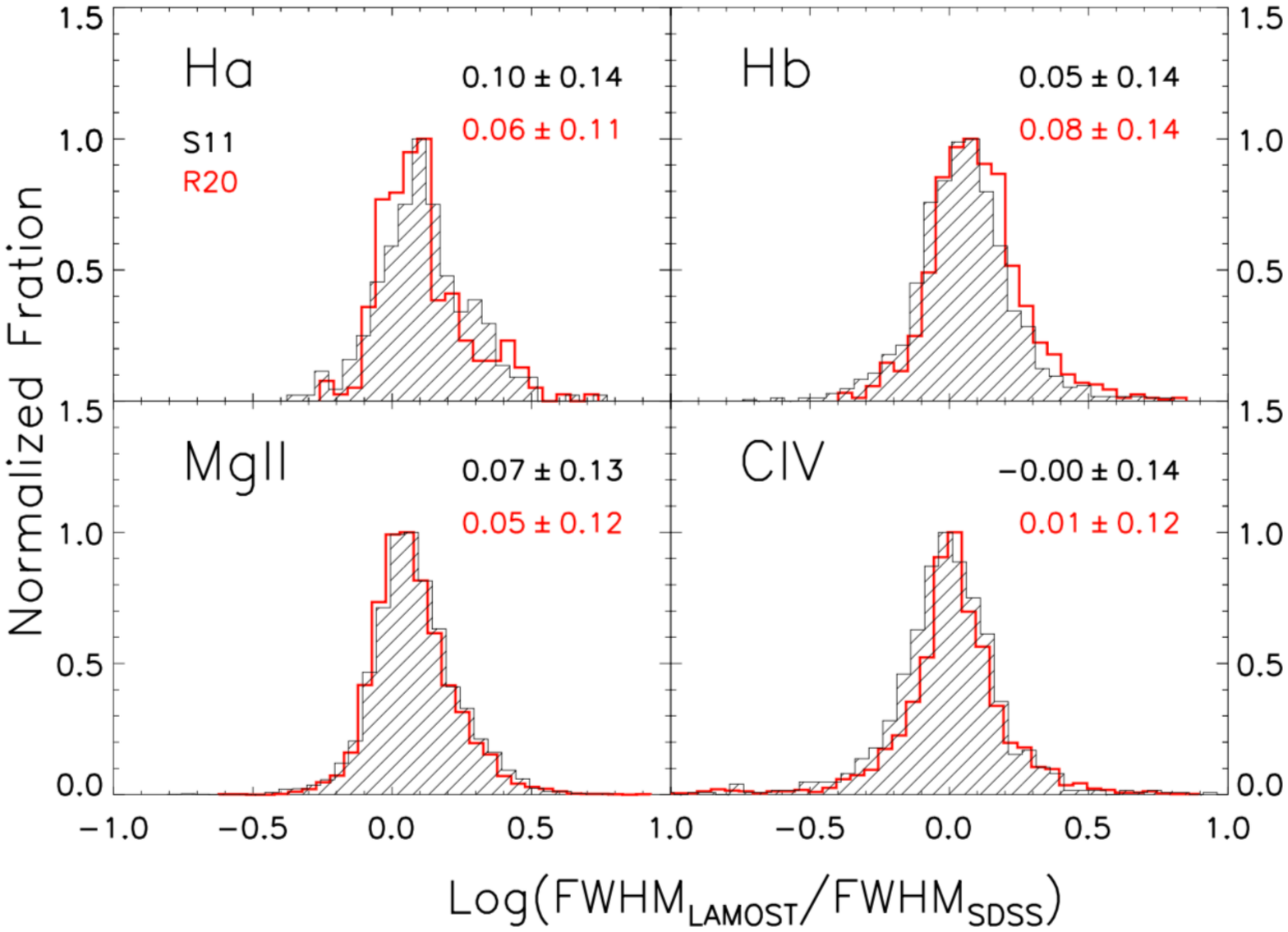}   
		\caption{Comparisons between the measurements of the FWHM values in this work and S11 (R20). We show the plot of $\rm log(FWHM_{LAMOST}/FWHM_{SDSS})$ for broad H$\alpha$ (upper left), broad H$\beta$ (upper right), broad Mg\,{\sc ii} (lower left) and whole C\,{\sc iv} (lower right). The mean ($\mu$) and dispersion ($\sigma$) of  each distribution are tabulated in corresponding plots. In this figure, only the emission lines with reliable fitting ({\tt LINE\_FLAG = 0}) are considered.} 
	\end{center}
	\label{fig:Com_FWHM} 
\end{figure}

\begin{figure}[!htb]
	\begin{center}	
		\includegraphics[page=1,scale=0.39]{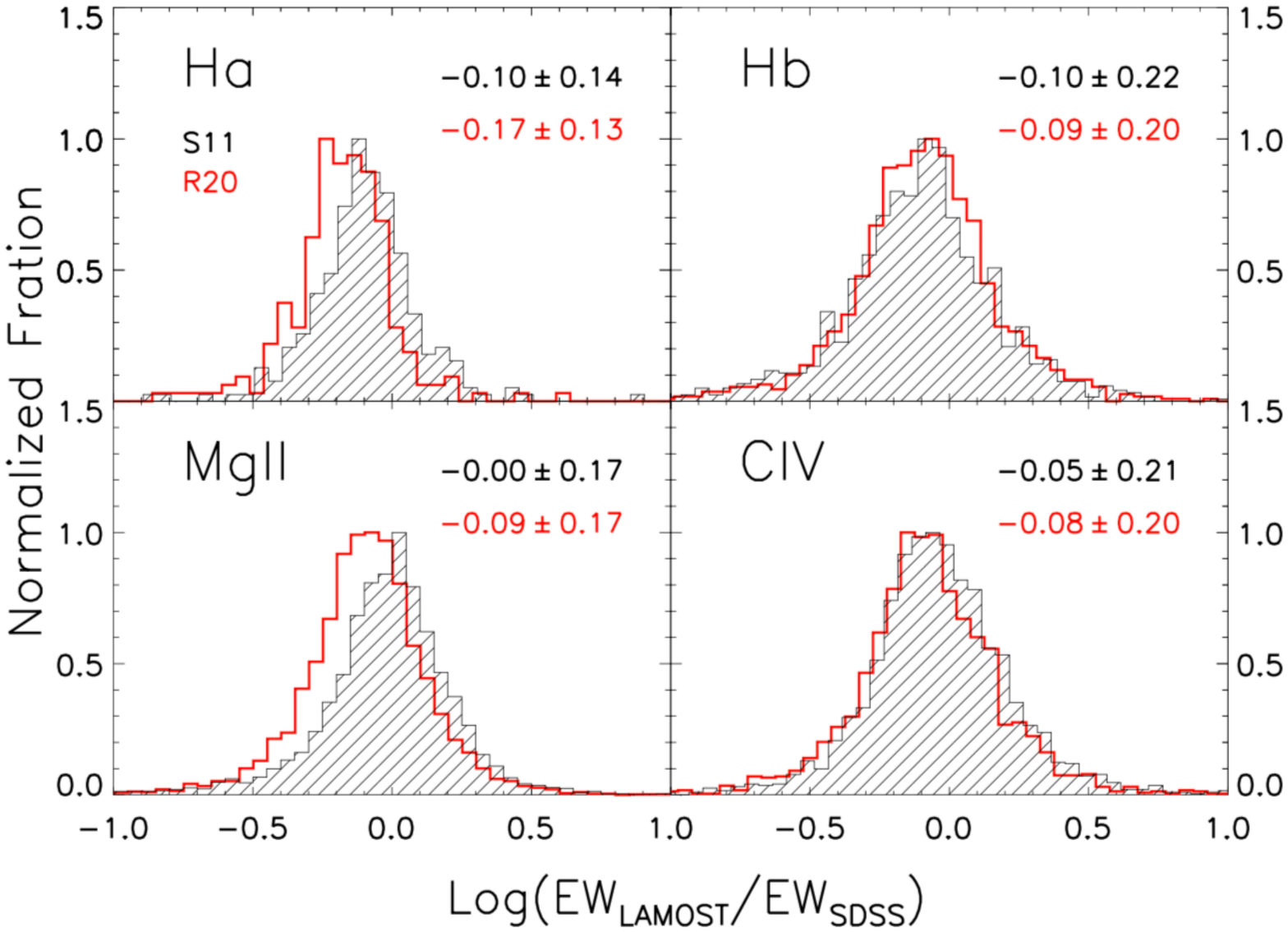}   
		\caption{Same as in Figure ~\ref{fig:Com_FWHM}, but for EW values.} 
	\end{center}
	\label{fig:Com_EW} 
\end{figure}

\begin{figure}[!htb]
	\begin{center}	
		\includegraphics[page=1,scale=0.39]{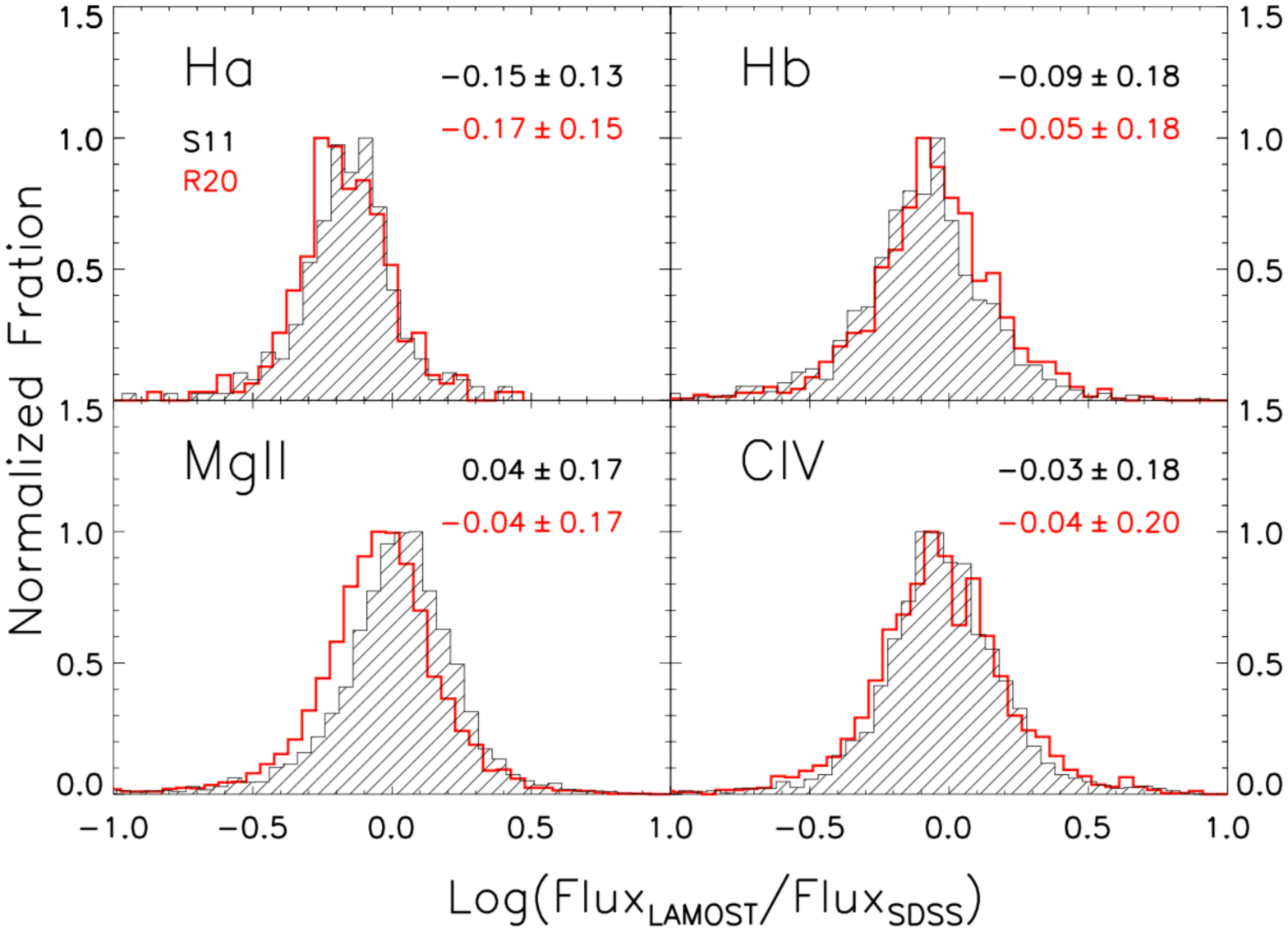}   
		\caption{Same as in Figure ~\ref{fig:Com_FWHM}, but for emission line flux.} 
	\end{center}
	\label{fig:Com_Flux} 
\end{figure}

There are 4964/6296 quasars in our catalog overlapping with the spectral fitting catalog of DR7Q/DR14Q (\citealt{2011ApJS..194...45S}, hereafter S11, and \citealt{2020ApJS..249...17R}, hereafter R20). To further justify the fitting results in this work, we compare the measured parameters for the common quasars between LAMOST DR6-9 and S11 (R20). The histograms in Figure ~\ref{fig:Com_FWHM} compare the logarithm FWHM values. In general, we find excellent agreement between the measurements. The mean ($\mu$) and standard deviation ($\sigma$) of the difference between this work and S11 (R20) is $0.10 \pm 0.14$ ($0.06 \pm 0.11$) for H$\alpha$, $0.05 \pm 0.14$ ($0.08 \pm 0.14$) for H$\beta$,  $0.07 \pm 0.13$ ($0.05 \pm 0.12$) for Mg\,{\sc ii}  and $0.00 \pm 0.14$ ($0.01 \pm 0.12$) for C\,{\sc iv} lines. The EW values in these two catalogs are also in agreement with each other (Figure ~\ref{fig:Com_EW} ). The $\mu$ and $\sigma$ between this work and S11 (R20) is $-0.10 \pm 0.14$ ($-0.17 \pm 0.13$) for H$\alpha$, $-0.10 \pm 0.22$ ($-0.09 \pm 0.20$) for H$\beta$,  $0.00 \pm 0.17$ ($-0.09 \pm 0.17$) for Mg\,{\sc ii}  and $-0.05 \pm 0.21$ ($-0.08 \pm 0.20$) for C\,{\sc iv} lines. As mentioned before, there are no emission flux information in the previous papers of the LAMOST quasars survey. In Figure ~\ref{fig:Com_Flux}, we show the comparison of the emission line flux measurements. The $\mu$ and $\sigma$ of the differences between this work and S11 (R20) is $-0.15 \pm 0.13$ ($-0.17 \pm 0.15$) for H$\alpha$, $-0.09 \pm 0.18$ ($-0.05 \pm 0.18$) for H$\beta$,  $0.04 \pm 0.17$ ($-0.04 \pm 0.17$) for Mg\,{\sc ii}  and $-0.03 \pm 0.18$ ($-0.04 \pm 0.20$) for C\,{\sc iv} lines. Similar to the FWHM and EW, the emission fluxes also show excellent agreement between the different measurements.

\begin{figure}[!htb]
	\begin{center}	
		\includegraphics[page=1,scale=0.39]{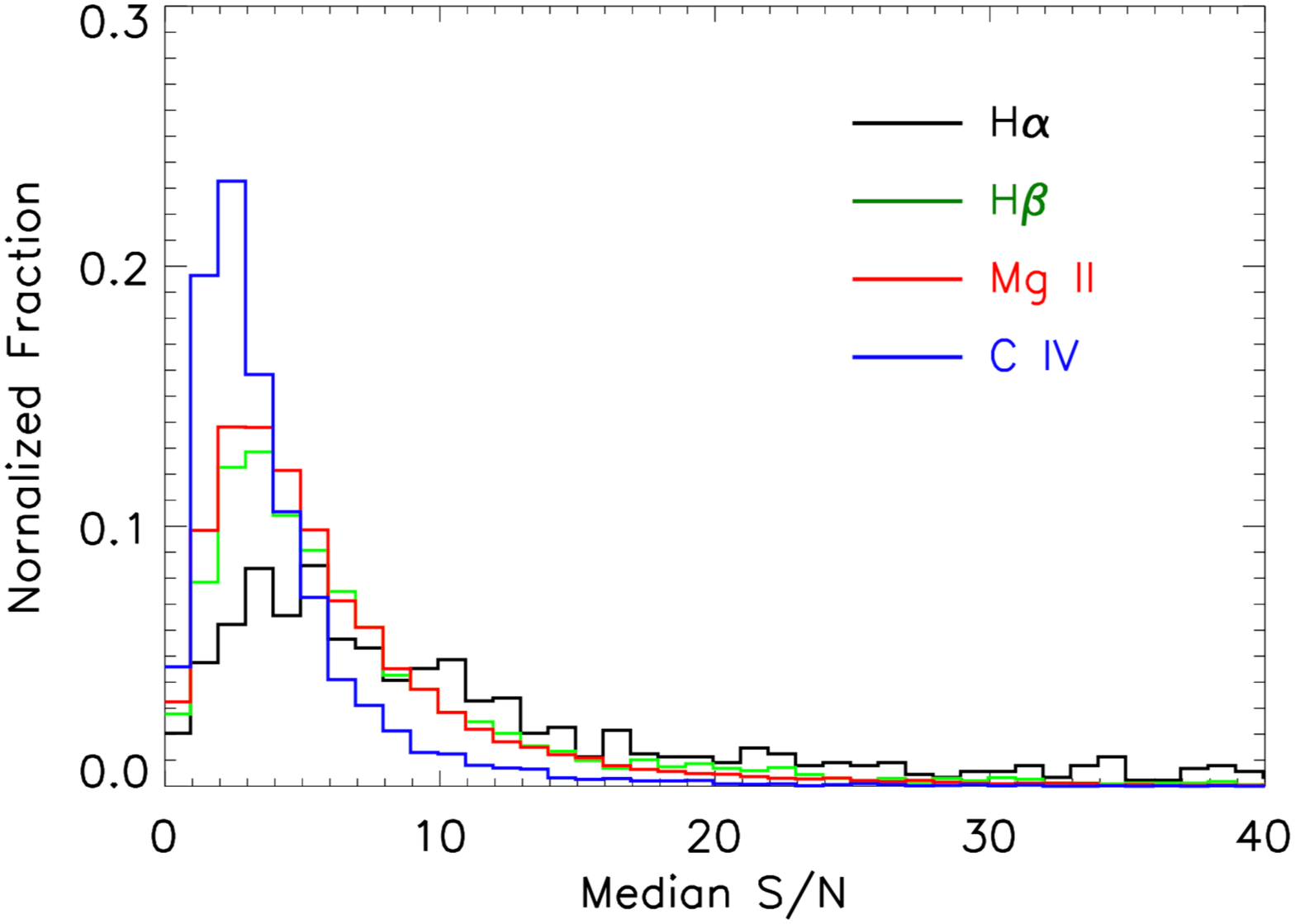}   
		\caption{The distributions of median S/N per pixel around the line-fitting region are plotted as normalized histogram. Only the emission lines with reliable fitting ({\tt LINE\_FLAG = 0}) are considered. } 
	\end{center}
	\label{fig:Fraction} 
\end{figure}

\begin{figure}[!htb]
	\begin{center}	
		\includegraphics[page=1,scale=0.39]{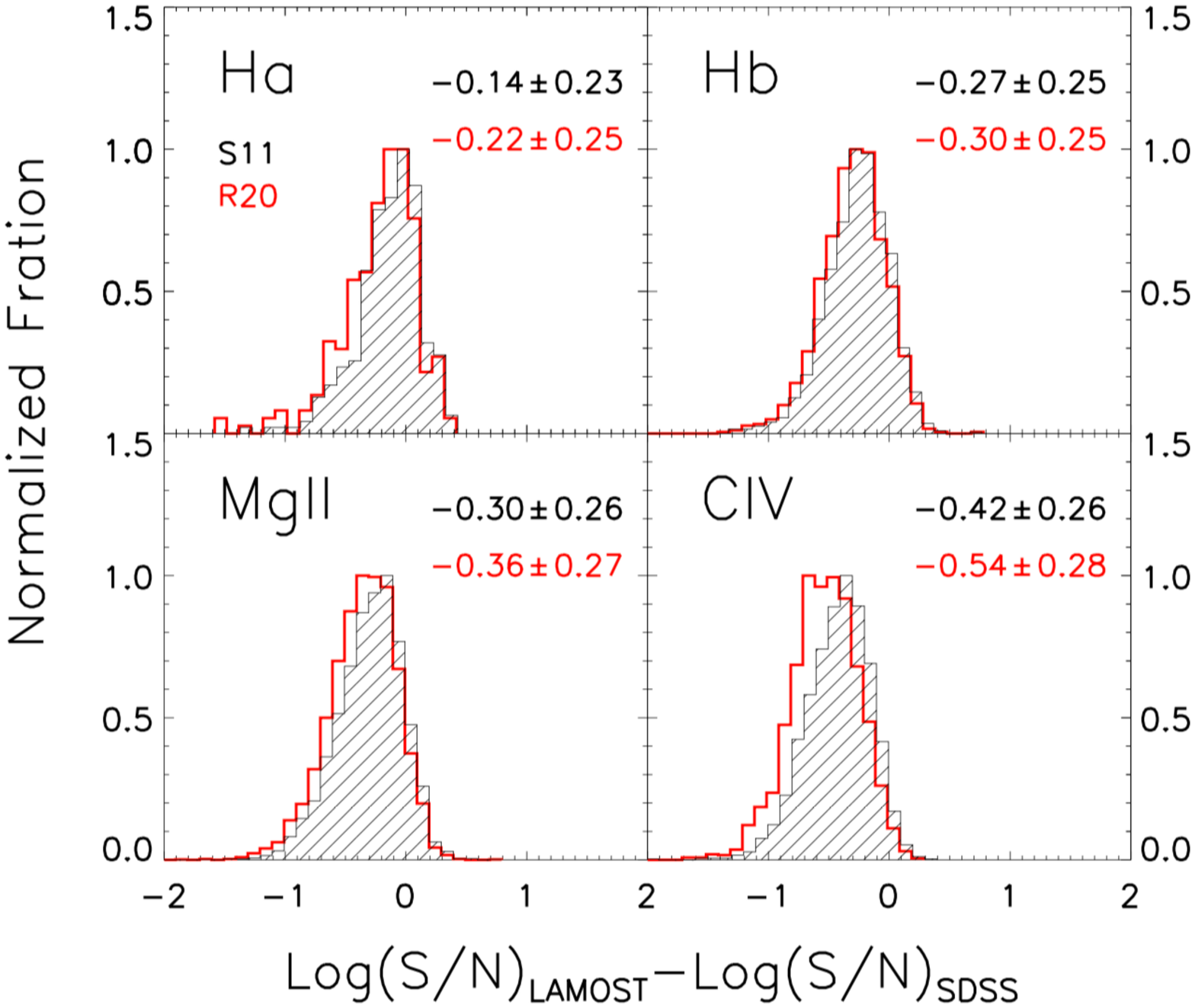}   
		\caption{The comparison of the median S/N per pixel in the line-fitting region between this work and S11  (R20). The mean ($\mu$) and dispersion ($\sigma$) of  each distribution are shown in corresponding plots. Only the emission lines with reliable fitting ({\tt LINE\_FLAG = 0}) are considered.} 
	\end{center}
	\label{fig:Com_snr} 
\end{figure}

In all cases, though a slight discrepancy between the different works is found, the measurements in this work are in agreement with those of SDSS. The differences may be caused by three main reasons: (1) Quasars usually show spectral variability, which can affect the measurements in different quasar catalogs. (2) The different S/N of SDSS and LAMOST spectra. As shown in Figure ~\ref{fig:Fraction}, the peak of the median S/N per pixel in line-fitting regions are all around or below S/N = 5. Figure ~\ref{fig:Com_snr} shows the comparison of the median S/N per pixel of the line-fitting regions between LAMOST DR6-9 and S11 (R20). It is clear that the LAMOST spectra have significantly lower S/N than those of SDSS spectra. (3) The different model used in the spectral fitting. For example, in the continuum fitting process, the host-galaxy subtraction is applied in R20, and there is an additional Balmer  continuum component in the pseudo-continuum. The Fe\,{\sc ii} template \citep{2001ApJS..134....1V} used in S11 is different from that in this work.  There are also some differences in the emission-line fitting process: we used the double Gaussians to model the broad component in H$\alpha$, H$\beta$, Mg\,{\sc ii} and C\,{\sc iv} emission lines, while in S11 or R20, multiple Gaussians (up to three) are used to fit each broad component. Moreover, in R20, there is no narrow component to model the C\,{\sc iv} emission lines.

The Monte Carlo (MC) approach is applied to estimate the uncertainty in each spectral fitting quantity.  The mock spectrum is produced by adding a Gaussian random noise (N(0, $\rm \sigma^{2}$), where the $\sigma$ represents the uncertainty in the spectrum per pixel) to the original spectrum. Then the spectral fitting is performed to the mock spectrum and the spectral quantities are estimated. The uncertainty of each quantity is then estimated as the standard deviation of the distribution given by 50 trials. 

\subsection{Virial Black hole Mass}

The monochromatic continuum luminosities at 1350 ($L_{1350}$), 3000 ($L_{3000}$), and 5100 ($L_{5100}$) $\rm \AA$ are calculated from the best-fit continuum $(f\rm_{bpl} + f\rm_{poly})$. 
By assuming the broad line region (BLR) is virialized, the $\rm M_{BH}$ can be estimated based on the single-epoch spectrum. The monochromatic continuum luminosity is used as a proxy of  the BLR radius, and the broad line width is used as a proxy of the virial velocity. The empirical scaling relation between the virial black hole mass and these two proxies are calibrated by AGN reverberation mapping. Here, the H$\beta$-based virial black hole masses are estimated using the relation \citep{2006ApJ...641..689V}:

\begin{equation}
\begin{split}
\log&M_{\rm BH}=\\
\log&
\left\{
\left[ 
\frac{\rm FWHM(H\beta)}{\rm km\,s^{-1}} 
\right]^{2} 
\left[
\frac{L_{5100}}{10^{44}\rm\,ergs\,s^{-1}}
\right]^{0.5} 
\right\}
+0.91,
\end{split}
\end{equation}
the Mg\,{\sc ii}-based virial black hole masses are estimated using the relation \citep{2009ApJ...707.1334W}:
\begin{equation}
\begin{split}
\log&M_{\rm BH}=\\
\log&
\left\{
\left[ 
\frac{\rm FWHM(Mg\,\text{\sc ii})}{\rm km\,s^{-1}} 
\right]^{1.51} 
\left[
\frac{L_{3000}}{10^{44}\rm\,ergs\,s^{-1}}
\right]^{0.5} 
\right\}
+2.60,
\end{split}
\end{equation}
and the C\,{\sc iv}-based virial black hole masses are estimated using the relation \citep{2006ApJ...641..689V}: 
\begin{equation}
\begin{split}
\log&M_{\rm BH}=\\
\log&
\left\{
\left[ 
\frac{\rm FWHM(C\,\text{\sc iv})}{\rm km\,s^{-1}} 
\right]^{2} 
\left[
\frac{L_{1350}}{10^{44}\rm\,ergs\,s^{-1}}
\right]^{0.53} 
\right\}
+0.66.
\end{split}
\end{equation}

\begin{figure}[!htb]
	\begin{center}	
		\includegraphics[page=1,scale=0.49]{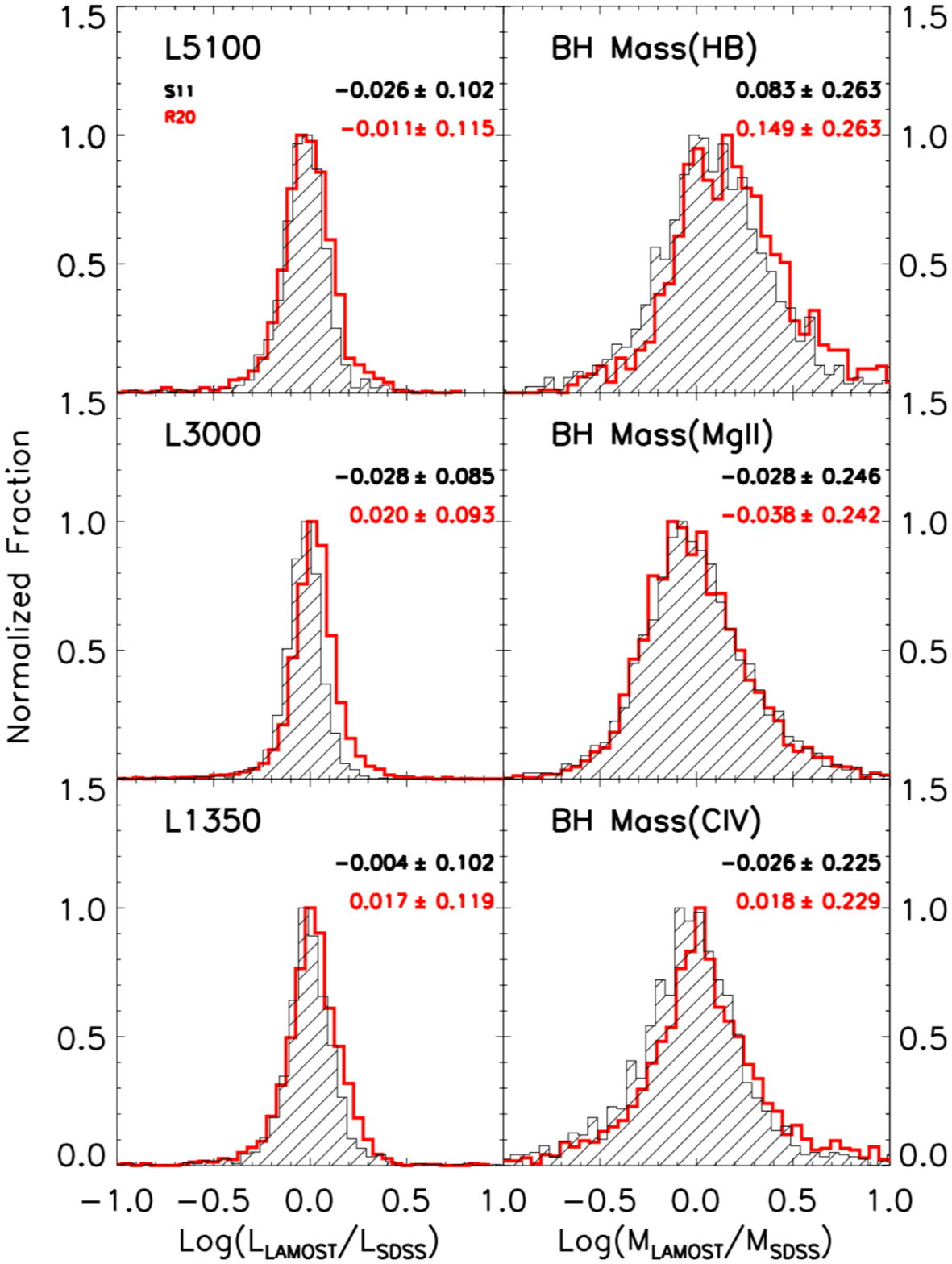}   
		\caption{The comparison of the monochromatic continuum luminosities ($L_{5100}$, $L_{1000}$, $L_{1300}$) and the estimated $\rm M_{BH}$  based on H$\beta$, Mg\,{\sc ii} and C\,{\sc iv} between this work and S11 (R20).} 
	\end{center}
	\label{fig:Com_Lcom_BHmass} 
\end{figure}

\begin{figure}[!htb]
	\begin{center}	
		\includegraphics[page=1,scale=0.49]{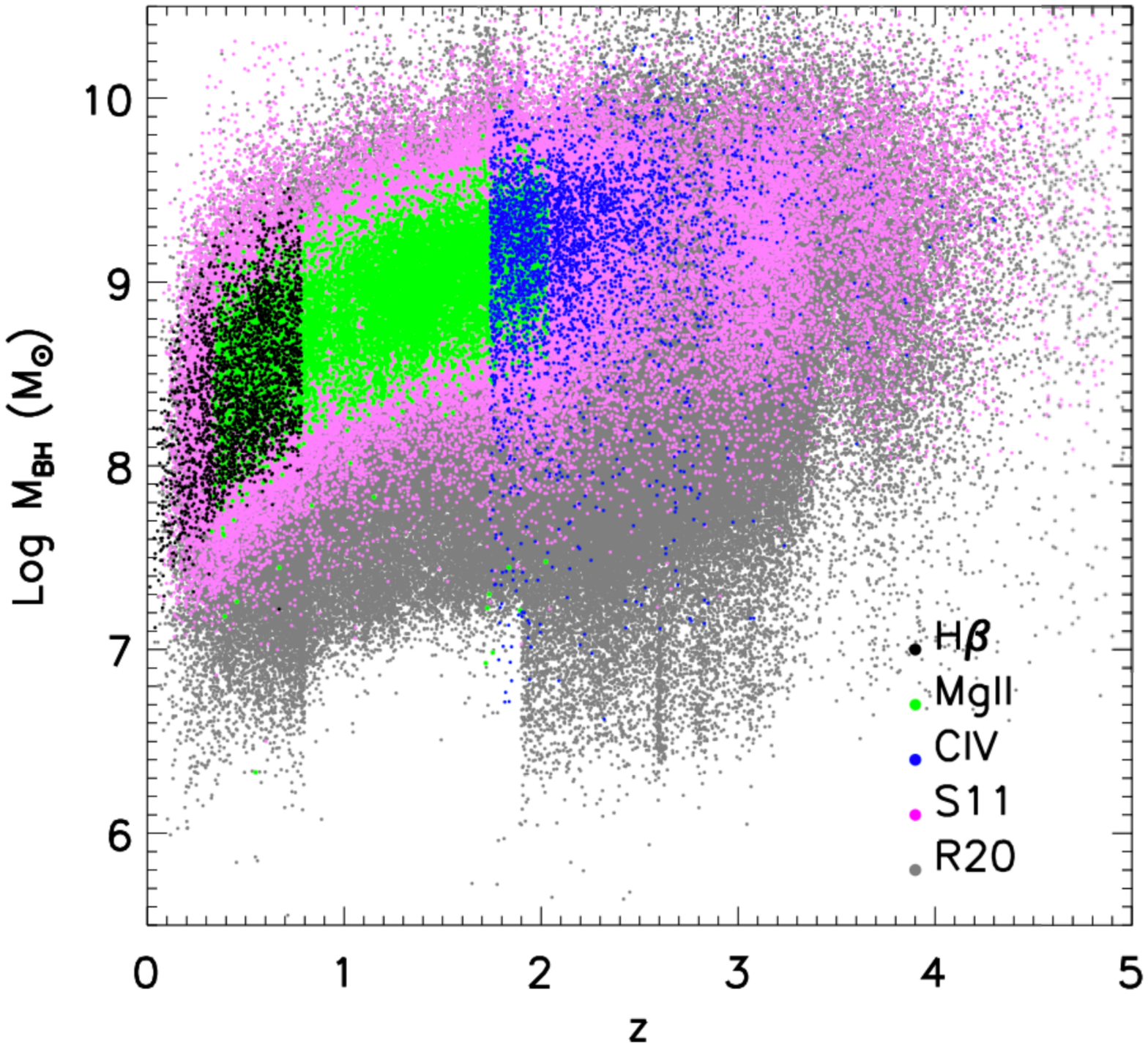}   
		\caption{The distribution of $\rm M_{BH}$ based on various broad emissions (H$\beta$, Mg\,{\sc ii} and C\,{\sc iv}) is plotted against the redshift. The quasars from S11 and R20 are represented by the pink and gray dots, respectively.} 
	\end{center}
	\label{fig:z_BHmass} 
\end{figure}

As mentioned before, the spectra are re-calibrated using the photometric data observed at epochs that are different from those of the LAMOST observations, which will introduce additional uncertainties, because of the variations of quasar luminosity, generally with magnitudes of 0.1-0.2 mag.  To justify this effect, we compare our continuum luminosities and $\rm M_{BH}$ measurements with those of S11 (R20) in Figure ~\ref{fig:Com_Lcom_BHmass}. In general, our estimates are in agreement with those of S11 (R20).  The deviations between the continuum luminosity measurements in this work and that of SDSS are generally with in 15\%.  If the variation of quasar luminosity is 0.1 (or 0.2) mag, the change of flux density estimated by the error transfer formula is $\rm \sim$  9.2\%(or 18.4\%). This means that the variation of quasar luminosity may dominate the error of the flux uncertainty for many quasars. Figure ~\ref{fig:z_BHmass} shows the distribution of the $\rm M_{BH}$ at different redshifts. Most quasars observed in SDSS DR7Q have low-to-moderate redshifts,  which is similar to the LAMOST survey.  While in SDSS DR14Q, compared with SDSS DR7Q, there are larger number of hight-redshift and low-luminosity quasars observed. Therefore it is apparent that the overall distribution of LAMOST quasars occupies the similar space as SDSS DR7Q, but has a relatively large discrepancy from SDSS DR14Q. The comparisons in both Figure ~\ref{fig:Com_Lcom_BHmass} and Figure ~\ref{fig:z_BHmass} prove that the flux re-calibration is mostly valid and the $\rm M_{BH}$ given in this work can be considered as a good approximation.

\section{Description of the Catalog}

We provide a compiled catalog for the quasars identified in LAMOST DR6-9 along with this paper. All measured quantities will be tabulated in the online catalog at LAMOST public website\footnote{https://nadc.china-vo.org/?locale=en}.
A summary of the parameters are listed in Table~\ref{tab:catalog2} and described as below.

\begin{itemize}[noitemsep]
	\item[1.] Unique spectra ID in LAMOST database. 
	\item[2.] Target Observation date. 
	\item[3.] LAMOST object designation: $\rm Jhhmmss.ss+ddmmss.s$ (J2000). 
	\item[4-5.] Right Ascension and Declination (in decimal degrees, J2000). 
	\item[6-9.] Spectroscopic observation information: 
	Local modified Julian date (LMJD), spectroscopic plan name (PlanID), spectrograph identification (spID), and spectroscopic fiber number (fiberID). 
	These four numbers are unique for each spectrum named in the format of 
	
	{spec$-$LMJD$-$planID\_spID$-$fiberID.fits}. 
	\item[10-11.] Redshift and its flag (ZWARNING) based on visual inspections. 1$=$not robust (eg., only one emission line available). 
	\item[12.] Target selection flag. 
	 SOURCE\_FLAG=``1'' indicates that the quasar was selected by its infrared-optical color, data-mining algorithms, multi-wavelength or other serendipitous algorithms. 
    SOURCE\_FLAG=``0'' means the object is not included in LAMOST quasar survey candidate sample but identified as quasar. 
	\item[13.] $\rm M_{i}$ $\rm (z=2)$: absolute i-band magnitude with K-corrected to $z=2$ following  \citet{2006AJ....131.2766R}. 
	\item[14.] Number of spectroscopic observations for the quasar. When there are more than one observations for the object, the line properties are obtained from only one of the observations in which the S/N is highest. 
	\item[15.] Median S/N per pixel in the continuum wavelength regions. 
	\item[16.] Flag of broad absorption features. BAL\_FLAG$=$1 indicates broad absorption features are present in Mg\,{\sc ii} and/or C\,{\sc iv}. 
	\item[17-46.] FLUX, FWHM, rest-frame EW, and their uncertainties for broad \ha, narrow \ha, [N\,{\sc ii}]$\lambda6584$ and [S\,{\sc ii}]$\lambda\lambda$6716,6731 emission lines. 
	\item[47-48.] Number of good pixels and median S/N per pixel for the spectrum in \ha\ region of rest-frame 6350-6800\,\AA. 
	\item[49.] Flag indicates reliability of the emission line fitting results in \ha\ region upon visual inspections. 0$=$acceptable; -1$=$unacceptable. 
	This value is set to be $-9999$ if \ha\ is not measured due to too few good pixels in the fitting region. 
	\item[50-67.] FLUX, FWHM, rest-frame EW, and their uncertainties for broad \hb, narrow \hb, [O\,{\sc iii}]$\lambda\lambda$5007 emission lines. 
	\item[68-69.] Number of good pixels and median S/N per pixel for the spectrum in \hb\ region of rest-frame 4600-5100\,\AA. 
	\item[70.] Flag indicates reliability of the emission line fitting results in \hb\ region upon visual inspections. 0$=$acceptable; -1$=$unacceptable. 
	This value is set to be $-9999$ if \hb\ is not measured due to too few good pixels in the fitting region. 
	\item[71-82.] FLUX, FWHM, rest-frame EW, and their uncertainties for the broad and narrow \mgii\ emission line. 
	\item[83-84.] Number of good pixels and median S/N per pixel for the spectrum in \mgii\ region of rest-frame 2700-2900\,\AA. 
	\item[85.] Flag indicates reliability of the emission line fitting results in \mgii\ region upon visual inspections. 0$=$acceptable; -1$=$unacceptable. 
	This value is set to be $-9999$ if \mgii\ is not measured due to too few good pixels in the fitting region. 
	\item[86-103.] FLUX, FWHM, rest-frame EW, and their uncertainties for the whole, broad and narrow \civ\ emission line. 
	\item[104-105.] Number of good pixels and median S/N per pixel for the spectrum in \civ\ region of rest-frame 1500-1700\,\AA. 
	\item[106.] Flag indicates reliability of the emission line fitting results in \civ\ region upon visual inspections. 0$=$acceptable; -1$=$unacceptable. 
	This value is set to be $-9999$ if \civ\ is not measured due to too few good pixels in the fitting region.
	\item[107.] Wavelength power-law index, $\alpha_\lambda$, from blueward of 4661\,\AA. 
	\item[108.] Wavelength power-law index, $\alpha_\lambda$, from redward of 4661\,\AA. 
	\item[109.] Rest-frame normalization parameter of optical Fe II.
	\item[110.] Rest-frame Gaussian FWHM of optical Fe II complex.
	\item[111.] Rest-frame wavelength shift of optical Fe II complex.
	\item[112.] Rest-frame normalization parameter of UV Fe II complex.
	\item[113.] Rest-frame Gaussian FWHM of UV Fe II complex.
	\item[114.] Rest-frame wavelength shift of UV Fe II complex.
	\item[115-120.] Monochromatic luminosities and their uncertanties at 1350, 3000 and 5100\,\AA. 
	\item[121-123.] Virial black hole masses (in $\rm M_{\odot}$) with calibrations of \hb, \mgii\ and \civ. 
	\item[124] Name of the quasar in SDSS quasar catalog. 
	The LAMOST DR6-9 quasar catalog was cross-correlated with the SDSS quasar catalog \citep[DR14,][]{2018A&A...613A..51P} using a matching radius of 3\arcsec. 
	\item[125.] Name of the object in second ROSAT all-sky survey point source catalog \citep[2RXS,][]{2016A&A...588A.103B}. 
	The LAMOST DR6-9 quasar catalog was cross-correlated with 2RXS using a matching radius of 30\arcsec. 
	The nearest point source in 2RXS was chosen. 
	\item[126-127.] The background corrected source counts in full band (0.1-2.4\,keV), and its error, from 2RXS. 
	\item[128.] The exposure time of the ROSAT measurement. 
	\item[129.] Angular separation between the LAMOST and 2RXS source positions. 
	\item[130.] Name of the object in XMM-Newton Serendipitous Source Catalog . The LAMOST DR6-9 quasar catalog was cross-correlated with XMM-Newton Serendipitous Source Catalog \citep[4XMM-DR11,][]{2020A&A...641A.136W} using a matching radius of 3\arcsec. 
	\item[131-132.] The mean full-band (0.2-12\,keV) flux, and its error, from 4XMM-DR11. 
	\item[133.] Angular separation between the LAMOST and 4XMM-DR11 source positions. 
	\item[134.] FIRST peak flux density at 20\,cm in units of mJy. 
	The LAMOST DR6-9 quasar catalog was cross-correlated with FIRST survey catalog using a matching radius of 5\arcsec. 
	\item[135.] Angular separation between LAMOST and FIRST source positions. 
	\item[136-143.] SDSS (or Pan-STARRS1) g, r, i, z PSF magnitudes without the correction for Galactic extinction, and their uncertainties.
	\item[144.] Flag of PSF magnitudes. `{\tt MAG\_FLAG$=$1}'  indicates the PSF magnitudes are given by SDSS, `{\tt MAG\_FLAG$=$0}' indicates the PSF magnitudes are give by Pan-STARRS1 and `{\tt MAG\_FLAG$=$-1}' indicates that the quasars don‘t have reliable photometric information. 
	\item[145-150.] WISE W1, W2, W3 instrumental profile-fit photometry magnitudes without the correction for Galactic extinction, and their uncertainties.
	\item[151-156.] WISE Y, J, K AperMag3 magnitudes without the correction for Galactic extinction, and their uncertainties. The AperMag3 magnitudes are the aperture corrected magnitudes measured by UKIDSS with $\rm 2^{\prime \prime}$ diameter, providing the most accurate estimate of the total magnitude \citep{2006MNRAS.372.1227D}.
\end{itemize}

\startlongtable
\begin{deluxetable*}{llll}
	\tabletypesize{\scriptsize} 
	\tablecaption{Catalog format for the quasars identified in LAMOST DR6-9 \label{tab:catalog2}}
	\tablehead{\colhead{Column} &  \colhead{Name} &  \colhead{Format} &  \colhead{Description}}  
	\startdata
	1   & ObsID  &  LONG  & Unique Spectra ID in LAMOST database \\
	2   & ObsDate & STRING  &  Target observation date \\
	3   & NAME  &  STRING  &  LAMOST designation hhmmss.ss+ddmmss (J2000) \\
	4   & RA            &  DOUBLE  &  Right ascension (R.A.) in decimal degrees (J2000) \\
	5   & DEC           &  DOUBLE  &  Declination (Decl.) in decimal degrees (J2000) \\
	6   & LMJD           &  LONG    &  Local Modified Julian Day of observation \\
	7   & PLANID       &  STRING  &  Spectroscopic plan identification \\
	8   & SPID          &  LONG  &  Spectrograph identification \\
	9   & FIBERID       &  LONG    &  Spectroscopic fiber number \\
	\hline
	10 & Z$\_$VI          &  DOUBLE  &  Redshift based on visual inspection \\
	11  & ZWARNING    &  LONG    &  ZWARNING flag based on visual inspection \\
	12	& SOURCE\_FLAG  & LONG   &  Flag of quasar candidate selection\\
	13  & MI$\_$Z2         &  DOUBLE  &  M$_{i} (z=2)$, K-corrected to $z=2$ following Richards et al. (2006) \\
	14  & NSPECOBS	              & LONG  &  Number of spectroscopic observations \\
	15  & SNR\_SPEC               &  DOUBLE &  Median S/N per pixel of the spectrum \\
	16  &  BAL\_FLAG        & LONG  &  Flag of broad absorption features \\
	\hline
	17	& FLUX\_BROAD\_HA	          &  DOUBLE	&  Flux of broad \ha\ in $\rm 10^{-17} erg$ $\rm cm^{-2}$ $\rm s^{-1}$\\
	18	& ERR\_FLUX\_BROAD\_HA	&  DOUBLE	&  Uncertainty in FLUX$_{\rm H\alpha,broad}$ \\
	19  & FWHM\_BROAD\_HA	      &  DOUBLE	&  FWHM of broad \ha\ in \kmps \\
	20  & ERR\_FWHM\_BROAD\_HA    &  DOUBLE	&  Uncertainty in FWHM$_{\rm H\alpha,broad}$ \\
	21	& EW\_BROAD\_HA	          &  DOUBLE	&  Rest-frame EW of broad \ha\ in \AA \\
	22	& ERR\_EW\_BROAD\_HA	&  DOUBLE	&  Uncertainty in EW$_{\rm H\alpha,broad}$ \\
	23  & FLUX\_NARROW\_HA	      &  DOUBLE	&  Flux of narrow \ha\ in $\rm 10^{-17} erg$ $\rm cm^{-2}$ $\rm s^{-1}$\\
	24  & ERR\_FLUX\_NARROW\_HA	  &  DOUBLE	&  Uncertainty in FLUX$_{\rm H\alpha,narrow}$ \\
	25	& FWHM\_NARROW\_HA	      &  DOUBLE	&  FWHM of narrow \ha\ in \kmps \\
	26	& ERR\_FWHM\_NARROW\_HA	  &  DOUBLE	&  Uncertainty in FWHM$_{\rm H\alpha,narrow}$ \\
	27  & EW\_NARROW\_HA	      &  DOUBLE	&  Rest-frame EW of narrow \ha\ in \AA \\
	28	& ERR\_EW\_NARROW\_HA	  &  DOUBLE	&  Uncertainty in EW$_{\rm H\alpha,narrow}$ \\
	29	& FLUX\_NII\_6584           &  DOUBLE	&  Flux of [N\,{\sc ii}]$\lambda$6584 in $\rm 10^{-17} erg$ $\rm cm^{-2}$ $\rm s^{-1}$\\
	30	& ERR\_FLUX\_NII\_6584	  &  DOUBLE	&  Uncertainty in FLUX$_{\rm[NII]6584}$ \\
	31	& FWHM\_NII\_6584           &  DOUBLE	&  FWHM of [N\,{\sc ii}]$\lambda$6584 in \kmps \\
	32 & ERR\_FWHM\_NII\_6584	  &  DOUBLE	&  Uncertainty in FWHM$_{\rm[NII]6584}$ \\
	33	& EW\_NII\_6584           &  DOUBLE	&  Rest-frame EW of [N\,{\sc ii}]$\lambda$6584 in \AA \\
	34	& ERR\_EW\_NII\_6584	  &  DOUBLE	&  Uncertainty in EW $_{\rm[NII]6584}$ \\
	35	& FLUX\_SII\_6716        &  DOUBLE	& Flux of [S\,{\sc ii}]$\lambda$6716 in $\rm 10^{-17} erg$ $\rm cm^{-2}$ $\rm s^{-1}$\\
	36	& ERR\_FLUX\_SII\_6716	  &  DOUBLE	&  Uncertainty in FLUX$_{\rm[SII]6716}$ \\
	37	& FWHM\_SII\_6716          &  DOUBLE	&  FWHM of [S\,{\sc ii}]$\lambda$6716 in \kmps \\
	38 & ERR\_FWHM\_SII\_6716	  &  DOUBLE	&  Uncertainty in FWHM$_{\rm[SII]6716}$ \\
	39	& EW\_SII\_6716        &  DOUBLE	&  Rest-frame EW of [S\,{\sc ii}]$\lambda$6716 in \AA \\
	40	& {ERR\_EW\_SII\_6716}	  &  DOUBLE	&  Uncertainty in EW $_{\rm[SII]6716}$ \\
	41	& FLUX\_SII\_6731         &  DOUBLE	&  Flux of [S\,{\sc ii}]$\lambda$6731 in $\rm 10^{-17} erg$ $\rm cm^{-2}$ $\rm s^{-1}$\\
	42	& ERR\_FLUX\_SII\_6731	  &  DOUBLE	&  Uncertainty in FLUX$_{\rm[SII]6731}$ \\
	43	& FWHM\_SII\_6731         &  DOUBLE	&  FWHM of [S\,{\sc ii}]$\lambda$6731 in \kmps  \\
	44	& ERR\_FWHM\_SII\_6731	  &  DOUBLE	&  Uncertainty in FWHM$_{\rm[SII]6731}$ \\
	45	& EW\_SII\_6731         &  DOUBLE	&  Rest-frame EW of [S\,{\sc ii}]$\lambda$6731 in \AA  \\
	46	& ERR\_EW\_SII\_6731	  &  DOUBLE	&  Uncertainty in EW$_{\rm[SII]6731}$ \\
	47	& LINE\_NPIX\_HA	      &  LONG	&  Number of good pixels for the rest-frame 6350-6800\,\AA  \\
	48	& LINE\_MED\_SN\_HA	      &  DOUBLE	&  Median S/N per pixel for the rest-frame 6350-6800\,\AA   \\
	49  & LINE\_FLAG\_HA        &  LONG   &  Flag for the quality in \ha\ fitting \\
	\hline
	50	& FLUX\_BROAD\_HB	         &  DOUBLE	&  Flux of broad \hb\ in $\rm 10^{-17} erg$ $\rm cm^{-2}$ $\rm s^{-1}$\\
	51	& ERR\_FLUX\_BROAD\_HB	  &  DOUBLE	&  Uncertainty in FLUX$_{\rm H\beta,broad}$ \\
	52  & FWHM\_BROAD\_HB	      &  DOUBLE	&  FWHM of broad \hb\ in \kmps \\
	53  & ERR\_FWHM\_BROAD\_HB    &  DOUBLE	&  Uncertainty in FWHM$_{\rm H\beta,broad}$ \\
	54	& EW\_BROAD\_HB	          &  DOUBLE	&  Rest-frame EW of broad \hb\ in \AA  \\
	55	& ERR\_EW\_BROAD\_HB	  &  DOUBLE	&  Uncertainty in EW$_{\rm H\beta,broad}$ \\
	56  & FLUX\_NARROW\_HB	      &  DOUBLE	&  Flux of narrow \hb\ in $\rm 10^{-17} erg$ $\rm cm^{-2}$ $\rm s^{-1}$\\
	57	& ERR\_FLUX\_NARROW\_HB	  &  DOUBLE	&  Uncertainty in FLUX$_{\rm H\beta,narrow}$ \\
	58	& FWHM\_NARROW\_HB      &  DOUBLE	&  FWHM of narrow \hb\ in \kmps \\
	59	& ERR\_FWHM\_NARROW\_HB	  &  DOUBLE	&  Uncertainty in FWHM$_{\rm H\beta,narrow}$ \\
	60  & EW\_NARROW\_HB	      &  DOUBLE	&  Rest-frame EW of narrow \hb\ in \AA \\
	61	& ERR\_EW\_NARROW\_HB	  &  DOUBLE	&  Uncertainty in EW$_{\rm H\beta,narrow}$ \\
	62	& FLUX\_OIII\_5007      &  DOUBLE	&  Flux of [O\,{\sc iii}]$\lambda$5007 in $\rm 10^{-17} erg$ $\rm cm^{-2}$ $\rm s^{-1}$\\
	63	& ERR\_FLUX\_OIII\_5007     &	 DOUBLE	&  Uncertainty in FLUX$_{\rm[OIII]5007}$ \\
	64	& FWHM\_OIII\_5007   &  DOUBLE	 &  FWHM of [O\,{\sc iii}]$\lambda$5007 in \kmps \\
	65	& ERR\_FWHM\_OIII\_5007  & DOUBLE & Uncertainty in FWHM$_{\rm[OIII]5007}$ \\
	66	& EW\_OIII\_5007	      &  DOUBLE	&  Rest-frame EW of [O\,{\sc iii}]$\lambda$5007 in \AA \\
	67	& ERR\_EW\_OIII\_5007     &	 DOUBLE	&  Uncertainty in EW$_{\rm[OIII]5007}$ \\
	68	& LINE\_NPIX\_HB	      &  LONG	&  Number of good pixels for the rest-frame 4600-5100\,\AA  \\
	69	& LINE\_MED\_SN\_HB	      &  DOUBLE	&  Median S/N per pixel for the rest-frame 4600-5100\,\AA  \\
	70  & LINE\_FLAG\_HB         &  LONG   &  Flag for the quality in \hb\ fitting\\
	\hline
	71	& FLUX\_BROAD\_MGII	      &  DOUBLE	&  Flux of the broad \mgii\ in $\rm 10^{-17} erg$ $\rm cm^{-2}$ $\rm s^{-1}$\\
	72	& ERR\_FLUX\_BROAD\_MGII	  &  DOUBLE	&  Uncertainty in FLUX$_{\rm MgII,broad}$ \\
	73	& FWHM\_BROAD\_MGII	      &  DOUBLE	&  FWHM of the broad \mgii\ in \kmps \\
	74	& ERR\_FWHM\_BROAD\_MGII  &  DOUBLE	&  Uncertainty in FWHM$_{\rm MgII,broad}$  \\
	75	& EW\_BROAD\_MGII	      &  DOUBLE	&  Rest-frame EW of the broad \mgii\ in \AA \\
	76	& ERR\_EW\_BROAD\_MGII	  &  DOUBLE	&  Uncertainty in EW$_{\rm MgII,broad}$ \\
	77	& FLUX\_NARROW\_MGII	      &  DOUBLE	&  Flux of the narrow \mgii\ in $\rm 10^{-17} erg$ $\rm cm^{-2}$ $\rm s^{-1}$\\
	78	& ERR\_FLUX\_NARROW\_MGII	  &  DOUBLE	&  Uncertainty in FLUX$_{\rm MgII,narrow}$ \\
	79  & FWHM\_NARROW\_MGII      &  DOUBLE &  FWHM of the narrow \mgii$\lambda$ in \kmps \\
	80  & ERR\_FWHM\_NARROW\_MGII &  DOUBLE &  Uncertainty in FWHM$_{\rm MgII,narrow}$ \\
	81	& EW\_NARROW\_MGII	      &  DOUBLE	&  Rest-frame EW of the narrow \mgii\ in \AA \\
	82	& ERR\_EW\_NARROW\_MGII	  &  DOUBLE	&  Uncertainty in EW$_{\rm MgII,narrow}$ \\
	83	& LINE\_NPIX\_MGII        &	 LONG	&  Number of good pixels for the rest-frame 2700-2900\,\AA  \\
	84	& LINE\_MED\_SN\_MGII	  &  DOUBLE	&  Median S/N per pixel for the rest-frame 2700-2900\,\AA   \\
	85  & LINE\_FLAG\_MGII        &  LONG   &  Flag for the quality in MgII fitting \\
	\hline
	86	& FLUX\_CIV	              &  DOUBLE &	Flux of the whole \civ\ in $\rm 10^{-17} erg$ $\rm cm^{-2}$ $\rm s^{-1}$\\
	87	& ERR\_FLUX\_CIV    &  DOUBLE	&   Uncertainty in Flux$_{\rm CIV,whole}$ \\
	88	& FWHM\_CIV	              &  DOUBLE &	FWHM of the whole \civ\ in \kmps \\
	89	& ERR\_FWHM\_CIV    &  DOUBLE	&   Uncertainty in FWHM$_{\rm CIV,whole}$ \\
	90	& EW\_CIV	              &  DOUBLE	&   Rest-frame EW of the whole \civ\ in \AA \\
	91	& ERR\_EW\_CIV	   &  DOUBLE	&   Uncertainty in EW$_{\rm CIV,whole}$ \\
	92	& FLUX\_BROAD\_CIV	              &  DOUBLE &	Flux of the broad \civ\ in $\rm 10^{-17} erg$ $\rm cm^{-2}$ $\rm s^{-1}$\\
	93	& ERR\_FLUX\_BROAD\_CIV    &  DOUBLE	&   Uncertainty in Flux$_{\rm CIV,broad}$ \\
	94	& FWHM\_BROAD\_CIV	      &  DOUBLE &	FWHM of the broad \civ\ in \kmps \\
	95	& ERR\_FWHM\_BROAD\_CIV  &  DOUBLE	&   Uncertainty in FWHM$_{\rm CIV,broad}$ \\
	96	& EW\_BROAD\_CIV	      &  DOUBLE	&   Rest-frame EW of the broad \civ\ in \AA \\
	97	& ERR\_EW\_BROAD\_CIV	  &  DOUBLE	&   Uncertainty in EW$_{\rm CIV,broad}$ \\
	98	& FLUX\_NARROW\_CIV	      &  DOUBLE	&   Flux of the narrow \civ\ in $\rm 10^{-17} erg$ $\rm cm^{-2}$ $\rm s^{-1}$\\
	99	& ERR\_FLUX\_NARROW\_CIV	  &  DOUBLE	&   Uncertainty in Flux$_{\rm CIV,narrow}$ \\
	100	& FWHM\_NARROW\_CIV	      &  DOUBLE &	FWHM of the narrow \civ\ in \kmps \\
	101	& ERR\_FWHM\_NARROW\_CIV  &  DOUBLE	&   Uncertainty in FWHM$_{\rm CIV,narrow}$ \\
	102 & EW\_NARROW\_CIV	      &  DOUBLE	&   Rest-frame EW of the narrow \civ\ in \AA \\
	103	& ERR\_EW\_NARROW\_CIV	  &  DOUBLE	&   Uncertainty in EW$_{\rm CIV,narrow}$ \\
	104   & LINE\_NPIX\_CIV	      &  LONG	&   Number of good pixels for the rest-frame 1500-1700\,\AA  \\
	105 	& LINE\_MED\_SN\_CIV	  &  DOUBLE	&   Median S/N per pixel for the rest-frame 1500-1700\,\AA   \\
	106  & LINE\_FLAG\_CIV         &  LONG   &  Flag for the quality in CIV fitting \\
	\hline
	107    & ALPHA\_LAMBDA\_1   &  DOUBLE &   Wavelength power-law index from blueward of 4661\,\AA \\
	108    & ALPHA\_LAMBDA\_2   &  DOUBLE &   Wavelength power-law index from redward of 4661\,\AA \\
	109    & Fe\_op\_norm   &  DOUBLE &   The normalization applied to the optical $\rm Fe_{II}$ template \\
	110    & Fe\_op\_shift  &  DOUBLE &   The Gaussian FWHM applied to the optical the $\rm Fe_{II}$ template \\
	111    & Fe\_op\_FWHM   &  DOUBLE &   The wavelength shift applied to the optical $\rm Fe_{II}$ template \\
	112    & Fe\_uv\_norm   &  DOUBLE &   The normalization applied to the ultraviolet $\rm Fe_{II}$ template \\
	113    & Fe\_uv\_shift   &  DOUBLE &   The Gaussian FWHM applied to the ultraviolet the $\rm Fe_{II}$ template \\
	114    & Fe\_uv\_FWHM   &  DOUBLE &   The wavelength shift applied to the ultraviolet $\rm Fe_{II}$ template \\
	115	& LOGL1350	              & DOUBLE	&   Monochromatic luminosity at 1350\,\AA\ in $\rm erg\,s^{-1}$ \\
	116	& ERR\_LOGL1350	              & DOUBLE	&   Uncertainty in $\rm log L_{1350}$\\
	117	& LOGL3000	              & DOUBLE	&   Monochromatic luminosity at 3000\,\AA\ in $\rm erg\,s^{-1}$ \\
	118	& ERR\_LOGL3000	              & DOUBLE	&   Uncertainty in $\rm log L_{3000}$\\ 
	119	& LOGL5100	              & DOUBLE	&   Monochromatic luminosity at 5100\,\AA\ in $\rm erg\,s^{-1}$ \\
	120	& ERR\_LOGL5100	              & DOUBLE	&   Uncertainty in $\rm log L_{5100}$\\
	121	& LOGBH\_HB            &	DOUBLE	&   Virial BH mass (M$_{\sun}$) based on \hb \\
	122	& LOGBH\_MgII          &	DOUBLE	&   Virial BH mass (M$_{\sun}$) based on \mgii \\
	123	& LOGBH\_CIV           &	DOUBLE	&   Virial BH mass (M$_{\sun}$) based on \civ \\
	\hline
	124	& SDSS\_NAME	& STRING & Name of the quasar in the SDSS quasar catalog \\
	125	& 2RXS\_NAME	& STRING &  Name of the object in the 2nd ROSAT all-sky survey point source catalog \\
	126	& 2RXS\_CTS 	& DOUBLE & Background corrected source counts in 0.1-2.4\,keV from 2RXS source catalog \\
	127	& 2RXS\_ECTS 	& DOUBLE & Error of the source counts from 2RXS source catalog \\
	128	& 2RXS\_EXPTIME & DOUBLE & Source exposure time from 2RXS source catalog \\
	129	& LM\_2RXS\_SEP & DOUBLE & LAMOST-2RXS separation in arcsec \\
	130	& 4XMM\_NAME & STRING & Name of the object in XMM-Newton Serendipitous Source Catalog \\
	131	& 4XMM\_FLUX 	& DOUBLE & Flux in 0.2-12.0\,keV band from 4XMM-DR11 (in erg\,s$^{-1}$\,cm$^{-2}$) \\
	132	& 4XMM\_FLUX\_ERR 	& DOUBLE & Error of the flux in 0.2-12.0\,keV band from 4XMM-DR11 (in erg\,s$^{-1}$\,cm$^{-2}$) \\
	133	& LM\_4XMM\_SEP & DOUBLE & LAMOST-4XMM separation in arcsec \\
	134	& FPEAK	& DOUBLE & FIRST peak flux density at 20 cm in mJy \\
	135	& LM\_FIRST\_SEP	& DOUBLE & LAMOST-FIRST separation in arcsec \\
	\hline
	136 & g\_mag & DOUBLE & SDSS (or Pan-STARRS1  PSF) g magnitudes \\
	137 & ERR\_g\_mag & DOUBLE & g PSF magnitude errors\\
	138 & r\_mag & DOUBLE & SDSS (or Pan-STARRS1 ) r PSF magnitudes\\
	139 & ERR\_r\_mag &  DOUBLE & r PSF magnitude errors \\
	140 & i\_mag & DOUBLE & SDSS (or Pan-STARRS1 ) i PSF magnitudes\\
	141 & ERR\_i\_mag & DOUBLE & i PSF magnitude errors\\
	142 & z\_mag & DOUBLE & SDSS (or Pan-STARRS1 ) z PSF magnitudes\\
	143& ERR\_z\_mag & DOUBLE & z PSF magnitude errors\\
	144 & MAG\_FLAG & LONG & Flag of PSF magnitude\\	
	145 & W1\_mag & DOUBLE & instrumental profile-fit photometry magnitudes, W1 band\\
	146 & ERR\_W1\_mag & DOUBLE & W1 magnitude errors\\
    147 & W2\_mag & DOUBLE & instrumental profile-fit photometry magnitudes, W2 band\\
	148 & ERR\_W2\_mag & DOUBLE & W2 magnitude errors\\
	149 & W3\_mag & DOUBLE & instrumental profile-fit photometry magnitudes, W3 band\\
	150 & ERR\_W3\_mag & DOUBLE & W3 magnitude errors\\	
	151 & Y\_mag & DOUBLE & Y AperMag3 magnitudes ($\rm 2^{\prime \prime}$ aperture diameter)\\
	152 & ERR\_Y\_mag & DOUBLE & Y magnitude errors\\
	153 & J\_mag & DOUBLE & J AperMag3 magnitudes ($\rm 2^{\prime \prime}$ aperture diameter)\\
    154 & ERR\_J\_mag & DOUBLE & J magnitude errors\\	
	155 & K\_mag & DOUBLE & K AperMag3 magnitudes ($\rm 2^{\prime \prime}$ aperture diameter)\\
    156 & ERR\_K\_mag & DOUBLE & K magnitude errors\\	
	\enddata
	\tablenotetext{}{{(This table is available in its entirety in FITS format.)}}
\end{deluxetable*}

\section{Summary and Discussion} \label{sec:summary}

In this work, we present the result of LAMOST Quasar Survey in the sixth, seventh, eighth and ninth data releases. There are in total 13,066 visually confirmed quasars. Among the identified quasars, 
6,381 were reported by the SDSS DR14 quasar catalog after our survey began, while the remaining 6685 are considered as newly discovered. 

We applied the emission line measurements of H$\alpha$, H$\beta$, Mg\,{\sc ii} and C\,{\sc iv} for each confirmed quasar. As the LAMOST spectra lack information of absolute flux calibration, we re-calibrate the spectra by fitting the SDSS/Pan-STARRS1 photometric data. The measured quantities are compiled into the quasar catalog which is available on-line.

\begin{figure*}[!htb]
	\begin{center}	
		\includegraphics[page=1,scale=0.8]{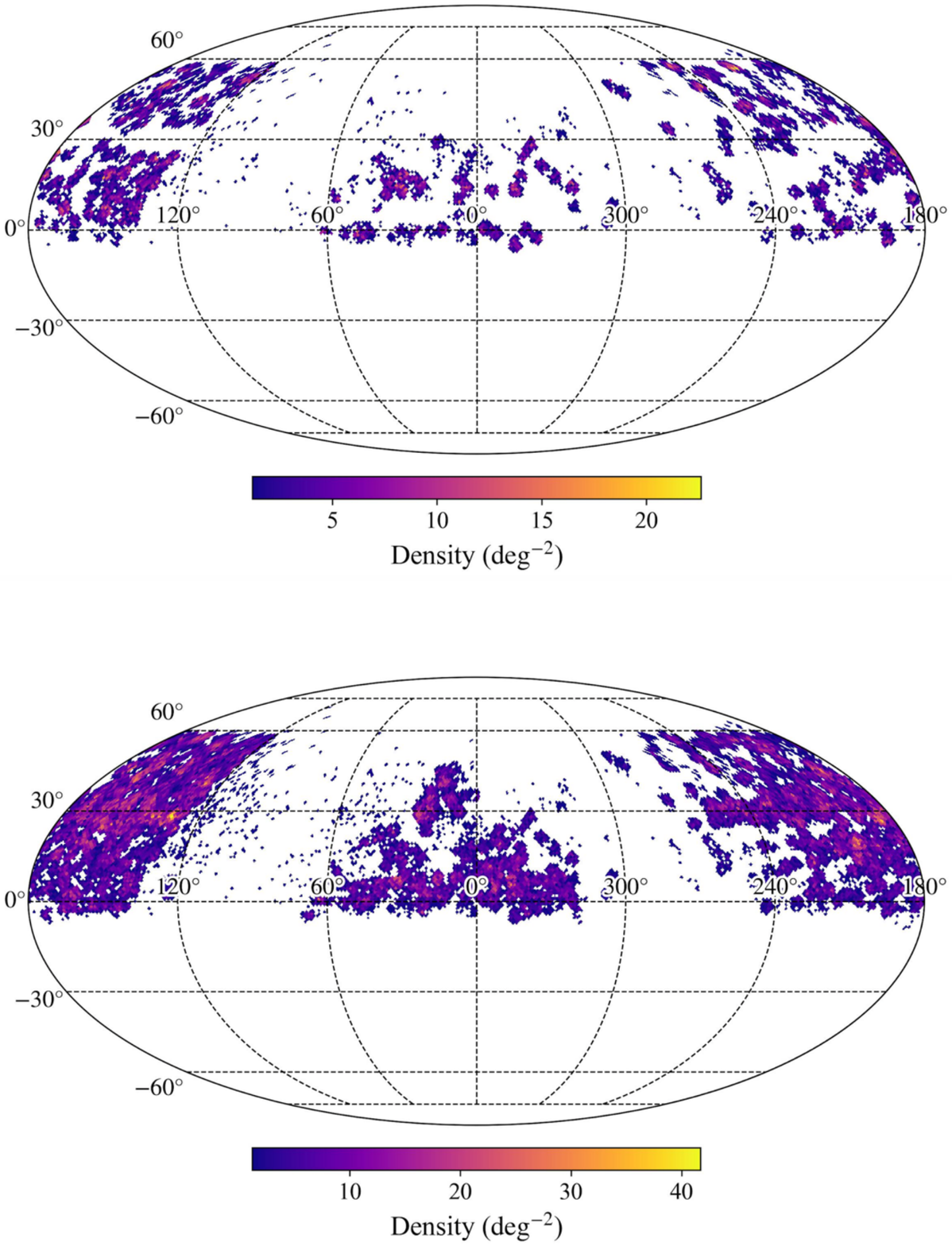}   
		\caption{The HEALPix sky distributions of the quasars identified in LAMOST DR6-9 (panel a) and DR1-9 (panel b) are shown in  Equatorial coordinates with the parameters $\rm N_{side=64}$ and area of 0.839 $\rm deg^{2}$ per pixel.} 
	\end{center}
	\label{fig:skyMap} 
\end{figure*}

After nine-year regular survey (\citealt{2016AJ....151...24A}, \citealt{2018AJ....155..189D}, \citealt{2019ApJS..240....6Y}, and this work),  there are in total 56,175 identified quasars in the LAMOST quasar survey, of which 31,866 are independently discovered by LAMOST. Among the identified quasars, 24,127 are newly discovered, and the remaining 32,048 are known ones that are reported by SDSS or Milliquas.(see Table~\ref{table:catalog3}). The sky distribution of LAMOST identified quasars is shown in Figure ~\ref{fig:skyMap}.


~\\

\begin{table}[]
	\centering
	\caption{The summary of the results of the LAMOST quasars survey up to now.}
	\scriptsize
	\begin{tabular}{lccccc}
		\hline 
		\hline 
		&Paper I&Paper II&Paper III&This Work&Total\\
		\hline 
		Total&3921&19935&19253&13066&56175\\
		Known&2741&11835&11091&6381&32048\\
		Independent&1180&12126&11458&7102&31866\\
		New&1180&8100&8162&6685&24127\\
		\hline 
	\end{tabular}
	\label{table:catalog3}
\end{table}

\begin{figure}[!htb]
	\begin{center}	
		\includegraphics[page=1,scale=0.49]{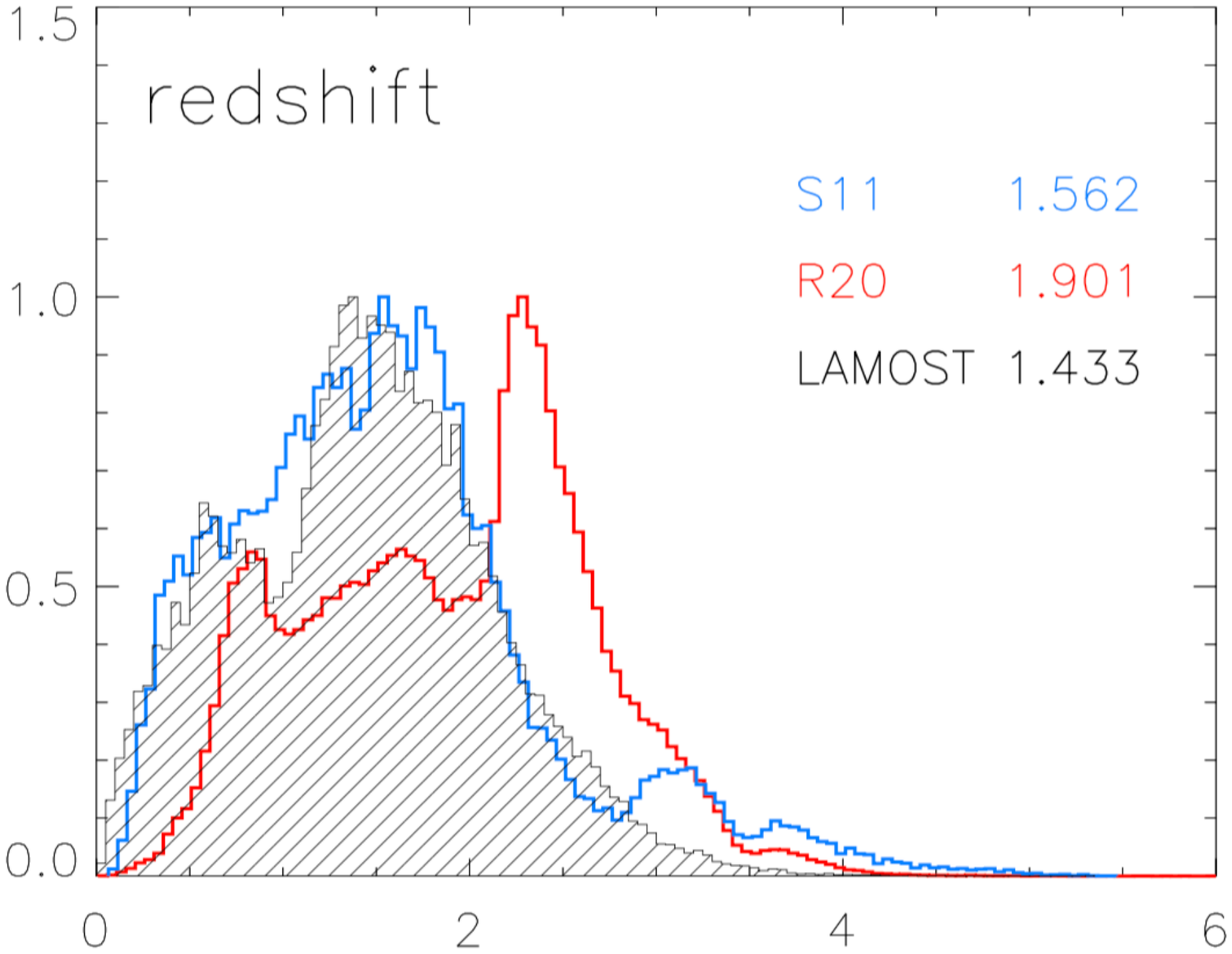}   
		\caption{The redshift distributions of LAMOST (black) and SDSS (blue for S11, and red for R20) samples. The mean redshifts are tabulated in the top-right corner.} 
	\end{center}
	\label{fig:redshift} 
\end{figure}

\begin{figure}[!htb]
	\begin{center}	
		\includegraphics[page=1,scale=0.49]{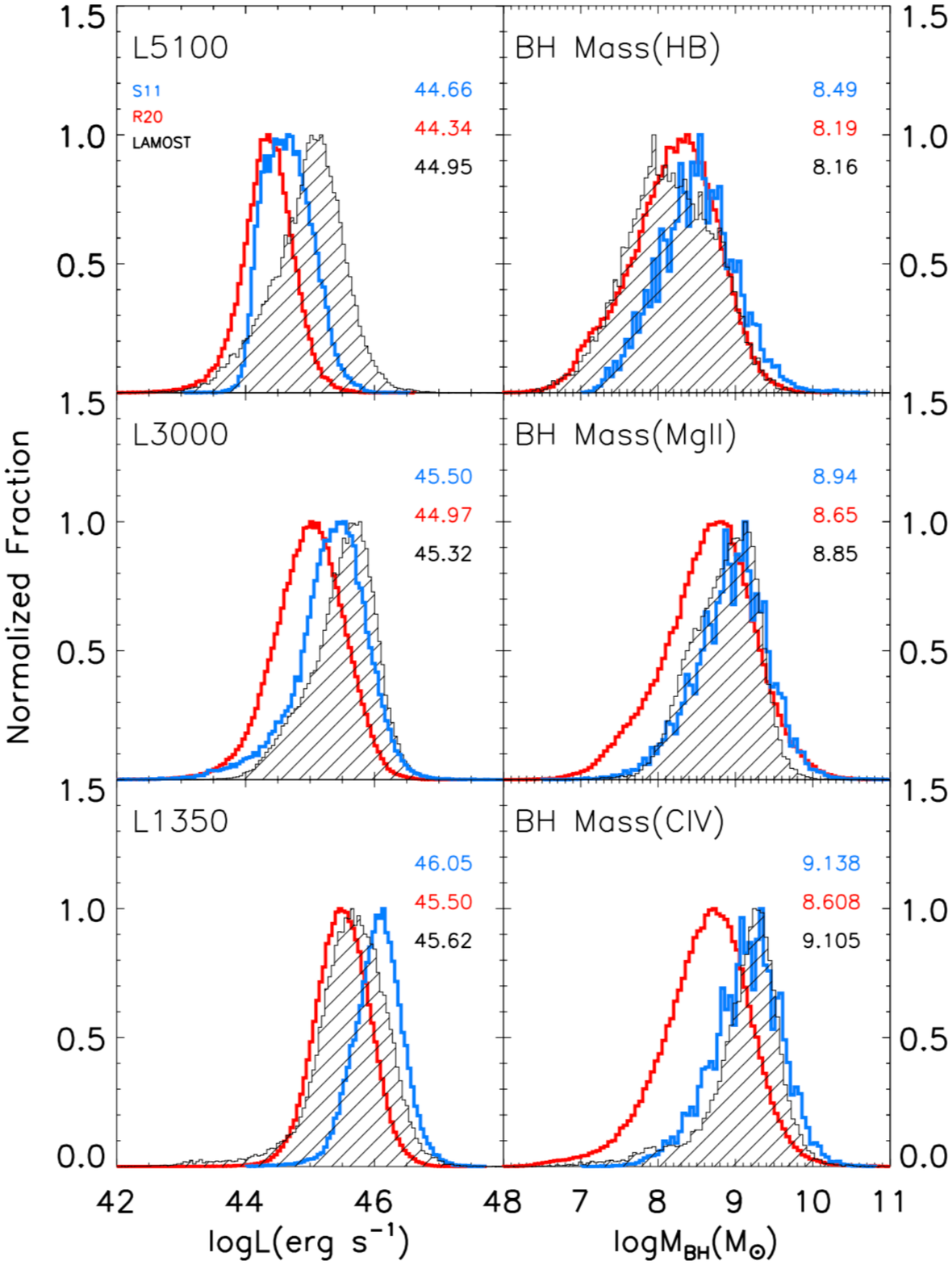}   
		\caption{The histograms of the monochromatic continuum luminosities ($L_{5100}$, $L_{1000}$, $L_{1300}$) and the estimated $\rm M_{BH}$  (based on H$\beta$, Mg\,{\sc ii} and C\,{\sc iv}) for LAMOST and S11 (R20) quasar sample. The mean value of each distribution is tabulated in the top-right corner.} 
	\end{center}
	\label{fig:L_M} 
\end{figure}

The basic properties of quasars identified in LAMOST DR1 to DR9 are compared with SDSS quasars. Figure ~\ref{fig:redshift} presents the redshift distribution of quasars for each sample. Generally, the redshift of LAMOST sample  is slightly lower than that of S11 and R20, but it is overall much more similar to S11 with only 0.13 smaller in the mean value.  The distributions of $\rm M_{BH}$ and continuum luminosities ($\rm L_{5100}$, $\rm L_{3000}$ and $\rm L_{1350}$) of LAMOST quasars are also compared with those of S11 and R20 in Figure ~\ref{fig:L_M}. As for the continuum luminosities of LAMOST sample, $\rm L_{5100}$ is higher than those of S11 and R20, $\rm L_{3000}$  is similar to those of S11, while $\rm L_{1350}$ is similar to those of R20. The distribution of $\rm H\beta$-based $\rm M_{BH}$ in LAMOST sample is similar to R20,  while the Mg\,{\sc ii}- and C\,{\sc iv}-based $\rm M_{BH}$ in LAMOST sample are similar to S11.  These distributions indicate that the quasars from LAMOST survey are brighter and have lower $\rm M_{BH}$ at lower redshift when compared with quasars from SDSS.

\begin{figure*}[!htb]
	\begin{center}	
		\includegraphics[page=1,scale=0.6]{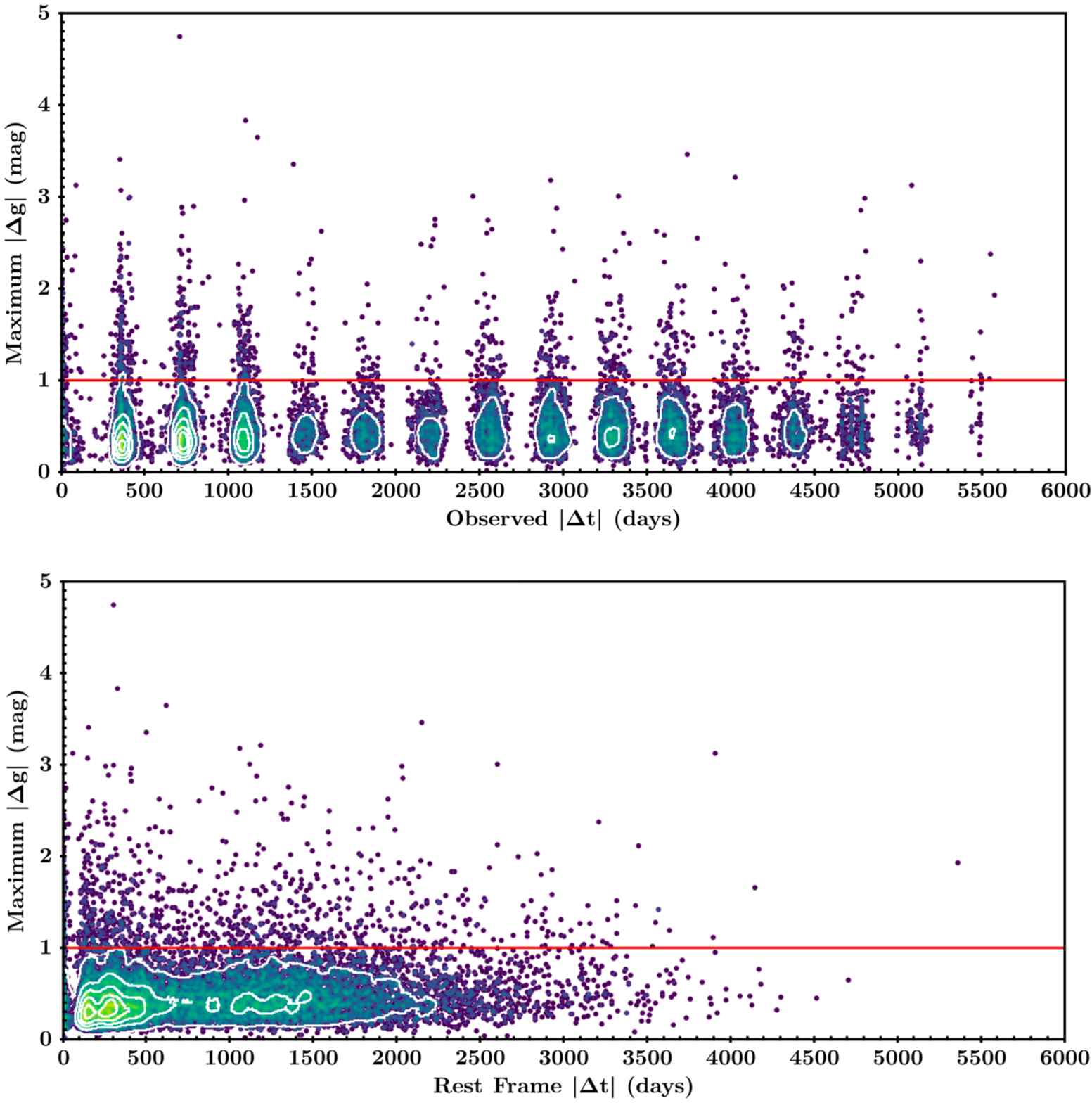}   
		\caption{Top panel: the distribution of the maximum magnitude difference based on SDSS and PS1 photometry $\rm |\Delta g|$ vs. time lag $\rm \Delta t$.  Bottom panel: The same distribution as top panel but the observed-frame lag is switching to the rest-frame lag by dividing ($\rm 1+z$).} 
	\end{center}
	\label{fig:vari} 
\end{figure*}

The optical variability of LAMOST quasars can be quantified by the maximum photometric difference between the g-band magnitude of SDSS and Pan-STARRS1 ($|\Delta g|_{max}$). To avoid the significant contamination from the poor photometry, we reject the photometric data with uncertainties $\rm \sigma _g \ge 0.15$ mag. Figure ~\ref{fig:vari} shows the distribution of $|\Delta g|_{max}$ versus time lag $|\Delta t|$ with contours.  The photometric variability of most quasars are within 1 magnitude, while some of them have at least a 1.0 mag change in their g-band light curves. Some previous works \citep{2016MNRAS.457..389M,2018ApJ...862..109Y} showed that the photometry varies following the spectral type transition, and the criterion $|\Delta g|_{max} > 1.0$ was used to select CL-AGN candidates. Some examples of CL-AGNs discovered in LAMOST quasar survey have been presented previously \citep{2018ApJ...862..109Y} and the follow-up works are on going.

\begin{figure*}[!htb]
	\begin{center}	
		\includegraphics[page=1,scale=0.71]{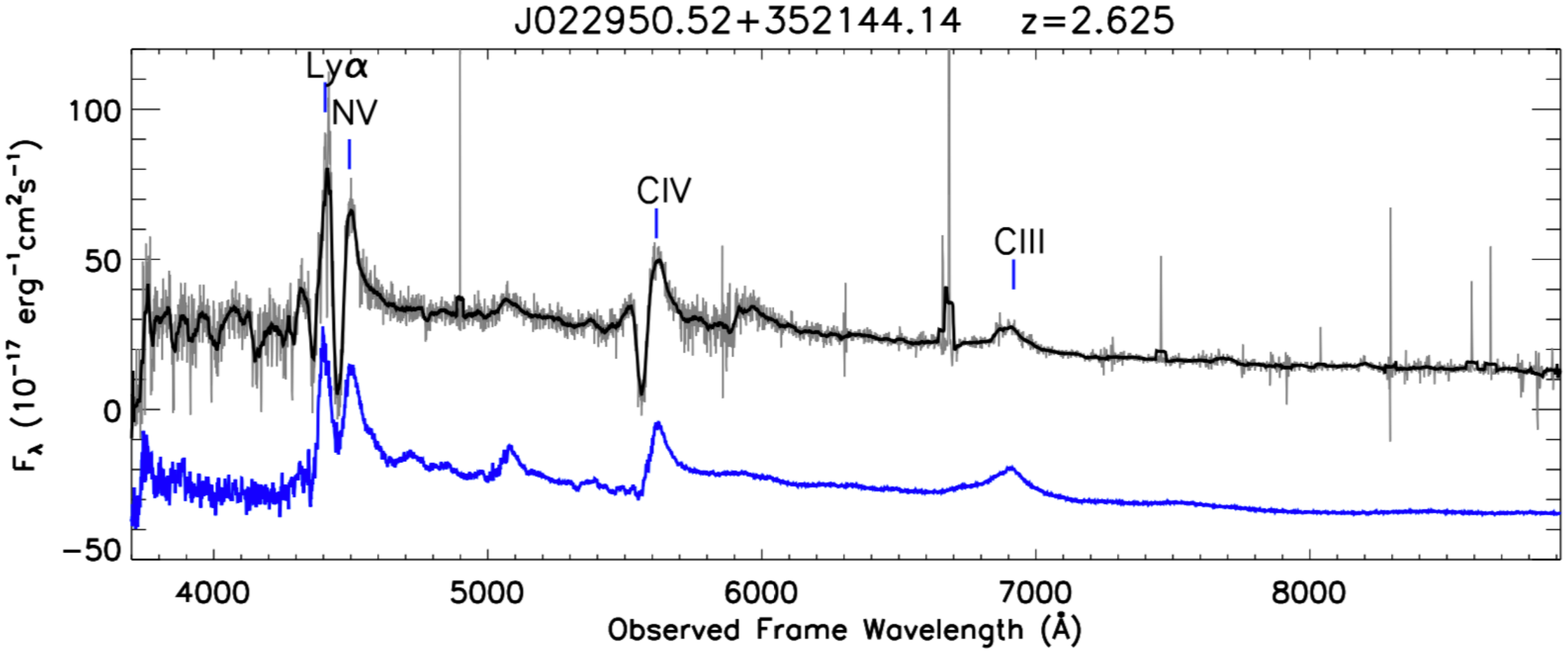}   
		\caption{An example for the $\rm Ly\alpha$ BAL quasar (black spectrum). The absorption features in the $\rm Ly \alpha$ and C\,{\sc iv} are obvious. The blue line is the BAL template spectra at the same redshift. There are no corrections for the redshift and Galactic extinction.} 
	\end{center}
	\label{fig:BAL} 
\end{figure*}

\begin{figure*}[!htb]
	\begin{center}	
		\includegraphics[page=1,scale=0.71]{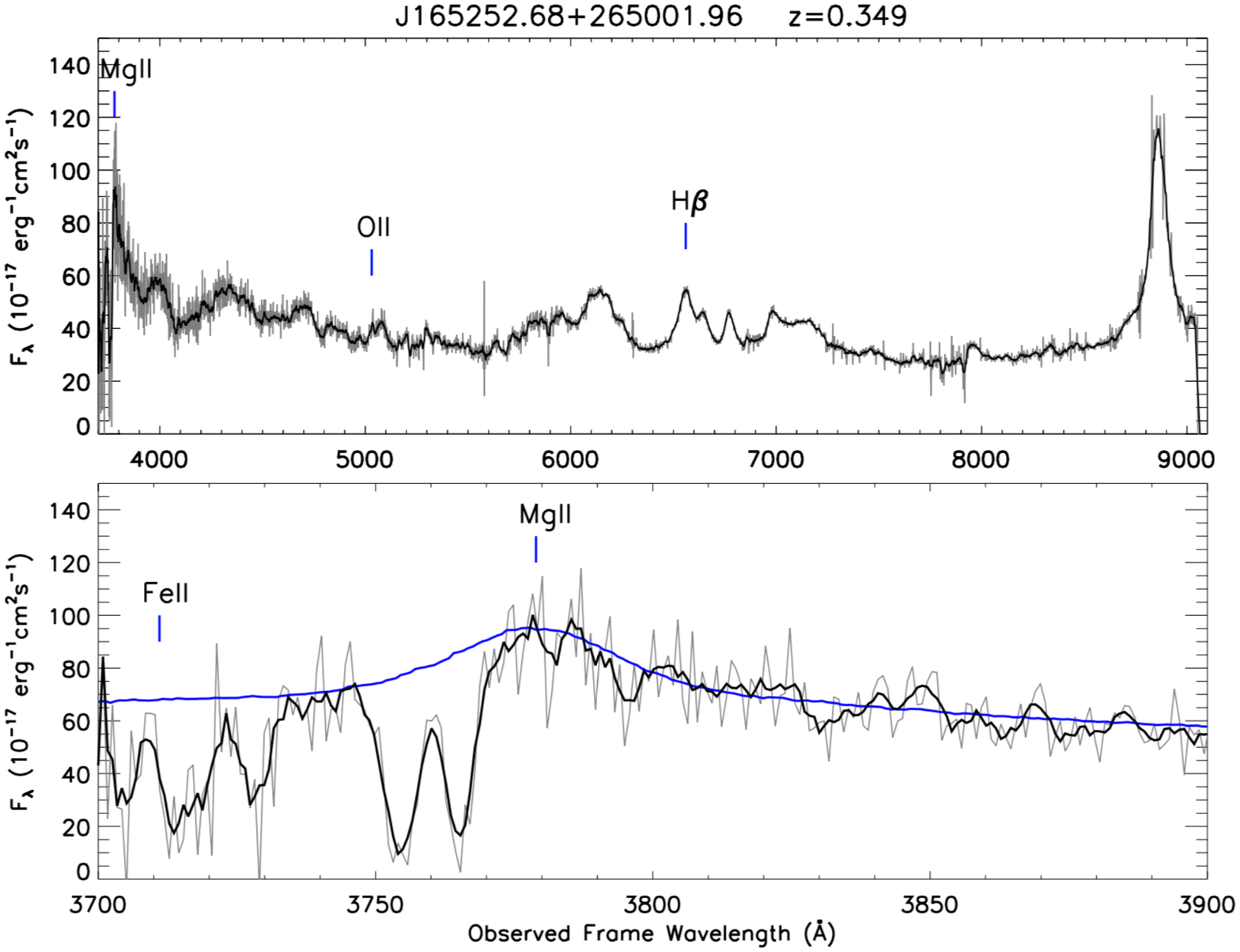}   
		\caption{An example for the LoBAL quasar. The top panel shows the whole spectrum, and the bottom panel is a zoom-in near the \mgii\ region. The blue line is the normal quasar template spectra at the same redshift. The \mgii\ absorption features is clear. There are no corrections for the redshift and Galactic extinction.} 
	\end{center}
	\label{fig:LoBAL} 
\end{figure*}

Except CL-AGNs, LAMOST quasars also include other kinds of interesting, unusual quasars. Figure ~\ref{fig:BAL} presents the spectrum of a $\rm Ly\alpha$ BAL quasar discovered during the visual inspection. Low-ionization BAL (LoBAL) quasars were also discovered in the survey.
	The spectrum of a LoBAL is ploted in Figure ~\ref{fig:LoBAL},  where the \mgii\  absorption features are obvious. 

The quasar catalog  provided by LAMOST is not only a great supplement to the low-to-moderate redshift quasars, but also a large database for investigating the quasar spectral variabilities and searching for unusual quasars.

It is worthwhile to launch the follow-up studies for discovery and investigating these interesting quasars in LAMOST quasar survey, such as the CL-AGN, BAL and LoBAL. Meanwhile, the LAMOST on going survey will extend its systematic searches to the quasars behind  the Galactic plane (GPQs) \citep{2021ApJS..254....6F,2022ApJS..261...32F}, which will provide more valuable data in the future.

\acknowledgments

This work is supported by 
the NSFC grants No.11721303, 11927804, 12133001 and the Ministry of Science and Technology of China under grant 2016YFA0400703.  
The author acknowledges the support by the China Postdoctoral Science Foundation (2021M690229), and the science research grant from the China Manned Space Project with No. CMS-CSST-2021-A06.

Guoshoujing Telescope (the Large Sky Area Multi-Object Fiber Spectroscopic Telescope LAMOST) is a National Major Scientific Project built by the Chinese Academy of Sciences. Funding for the project has been provided by the National Development and Reform Commission. LAMOST is operated and managed by the National Astronomical Observatories, Chinese Academy of Sciences.

This publication makes use of data products from the Sloan Digital Sky Survey. 
Funding for the Sloan Digital Sky Survey IV has been provided by the Alfred P. Sloan Foundation, the U.S. Department of Energy Office of Science, and the Participating Institutions. SDSS-IV acknowledges
support and resources from the Center for High-Performance Computing at
the University of Utah. The SDSS web site is www.sdss.org.

This work has made use of SDSS spectroscopic data. Funding for SDSS-III has been provided by the Alfred P. Sloan Foundation, the Participating Institutions, the National Science Foundation, and the U.S. Department of Energy Office of Science. The SDSS-III website is http://www.sdss3.org/. SDSS-III is managed by the Astrophysical Research Consortium for the Participating Institutions of the SDSS-III Collaboration, including 
the University of Arizona, the Brazilian Participation Group, Brookhaven National Laboratory, Carnegie Mellon University, University of Florida, the French Participation Group, the German Participation Group, Harvard University, the Instituto de Astrofisica de Canarias, the Michigan State/Notre Dame/ JINA Participation Group, Johns Hopkins University, Lawr- ence Berkeley National Laboratory, Max Planck Institute for Astrophysics, Max Planck Institute for Extraterrestrial Physics, New Mexico State University, New York University, Ohio State University, Pennsylvania State University, University of Portsmouth, Princeton University, the Spanish Participation Group, University of Tokyo, University of Utah, Vanderbilt University, University of Virginia, University of Washington, and Yale University.

The Pan-STARRS1 Surveys (PS1) and the PS1 public science archive have been made possible through contributions by the Institute for Astronomy, the University of Hawaii, the Pan-STARRS Project Office, the Max-Planck Society and its participating institutes, the Max Planck Institute for Astronomy, Heidelberg and the Max Planck Institute for Extraterrestrial Physics, Garching, The Johns Hopkins University, Durham University, the University of Edinburgh, the Queen's University Belfast, the Harvard-Smithsonian Center for Astrophysics, the Las Cumbres Observatory Global Telescope Network Incorporated, the National Central University of Taiwan, the Space Telescope Science Institute, the National Aeronautics and Space Administration under Grant No. NNX08AR22G issued through the Planetary Science Division of the NASA Science Mission Directorate, the National Science Foundation Grant No. AST–1238877, the University of Maryland, Eotvos Lorand University (ELTE), the Los Alamos National Laboratory, and the Gordon and Betty Moore Foundation.

\end{document}